\documentclass[manuscript]{aastex} 
\begin{document}
\title{Quasar Selection Based on Photometric Variability}
\author{C.~L.~MacLeod\altaffilmark{1}, 
  K.~Brooks\altaffilmark{1}, 
  \v{Z}.~Ivezi\'{c}\altaffilmark{1}, 
  C.~S.~Kochanek\altaffilmark{2,3}, 
  R.~Gibson\altaffilmark{1}, 
  A.~Meisner\altaffilmark{1},
  S.~Koz{\l}owski\altaffilmark{2,4},
  B.~Sesar\altaffilmark{1}, 
  A.~C.~Becker\altaffilmark{1}, 
  W.~H.~de Vries\altaffilmark{5}}

\altaffiltext{1}{Department of Astronomy, University of
  Washington, Box 351580, Seattle, WA 98195 (cmacleod@astro.washington.edu)} 
\altaffiltext{2}{Department of Astronomy, The Ohio State University, 140 West
  18th Avenue, Columbus, OH 43210} 
\altaffiltext{3}{The Center for Cosmology and Astroparticle Physics, The Ohio
  State University, 191 West Woodruff Avenue, Columbus, OH 43210} 
\altaffiltext{4}{Warsaw University Observatory, Al.\ Ujazdowskie 4,
00-478 Warsaw, Poland}
\altaffiltext{5}{University of California, One Shields Ave,
  Davis, CA 95616}

\begin{abstract}
We develop a method for separating quasars from other variable point
sources using SDSS Stripe 82 light curve data for $\sim$10,000
variable objects. To statistically describe quasar variability, we use
a damped random walk model parametrized by a damping time scale,
$\tau$, and an asymptotic amplitude (structure function),
$\rm{SF}_{\infty}$. With the aid of an SDSS spectroscopically confirmed
quasar sample, we demonstrate that variability selection in typical
extragalactic fields with low stellar density can deliver complete
samples with reasonable purity (or efficiency, $E$). Compared to a
selection method based solely on the slope of the structure function,
the inclusion of the $\tau$ information boosts $E$ from 60\% to 
75\% while maintaining a highly complete sample (98\%) even in the 
absence of color information. For a completeness of $C=90$\%, $E$ is
boosted from 80\% to 85\%. Conversely, $C$ improves from 90\% to 97\%
while maintaining $E=80$\% when imposing a lower limit on $\tau$. With  
the aid of color selection, the purity can be further boosted to 96\%,  
with $C= 93$\%.  Hence, selection methods based on variability will
play an important role in the selection of quasars with data provided
by upcoming large sky surveys, such as Pan-STARRS and the Large Synoptic Survey
Telescope (LSST). For a typical (simulated) LSST cadence over 10 years
and a photometric accuracy of 0.03 mag (achieved at $i \approx 22$),
$C$ is expected to be 88\% for a simple sample selection criterion of
$\tau>100$~days. In summary, given an adequate survey cadence,
photometric variability provides an even better method than color
selection for separating quasars from stars.

\end{abstract}

\section{Introduction}
The selection of large samples of quasars, or active galactic nuclei (AGN), is a valuable tool in many
areas of study, including galaxy evolution and black hole growth
(Kollmeier et al.~2006), absorbing systems in the intergalactic medium
(Hennawi \& Prochaska 2007; McDonald \& Eisenstein 2007), and  
determining the large scale structure of the universe (Ross et al.~2009). 
Extending these studies to significantly fainter quasars at any given 
redshift is a major goal for the next-generation Large Synoptic Survey 
Telescope (LSST; Ivezi\'{c} et al.~2008). LSST will probe the 
luminosity function to $M_B=-23$ at $z\sim 2$, breaking the luminosity-redshift 
degeneracy that arises in current flux-limited samples such as the Sloan 
Digital Sky Survey (SDSS), which can only detect $M_B>-23$ quasars at $z<0.5$. 
Also, our understanding of AGN fueling mechanisms and the detailed 
physics of accretion disks is far from complete (for a review, see
McHardy~2010).  Therefore, larger and more representative samples will
only provide more potential for discovery.  

Unfortunately, the resemblance of unresolved quasars (i.e., Type I
broad line AGN) to stars in photometric surveys poses a difficulty in 
their identification. Quasars can be reliably identified
by their spectra, and the SDSS provides a
spectroscopically complete survey to $i<19$ for the quasar region in
SDSS color space (Richards et al.~2002) which includes $\sim$105,000
quasars (Schneider et al.~2010). However, it is becoming increasingly important
to identify quasars in the absence of spectra, particularly since
complete spectroscopy will be too expensive in upcoming large-area sky
surveys.  Previously, quasar selection in the absence of spectroscopy
has best been achieved using $u-g$ and $g-r$ colors (e.g., Richards et
al.~2002; Jiang et al.~2006; Richards et al.~2009). This technique
requires multiple filters and results in samples with
efficiencies\footnote{\footnotesize Throughout this paper,
  ``efficiency'' ($E$) is equivalent to ``purity'' ($p$) in Schmidt et
  al.~(2010).} 
as low as 73\%, and there is a redshift regime near $z\simeq 3$ where 
optical color selection largely fails. The upcoming large-area sky
surveys, such as LSST and Pan-STARRS (Kaiser et al.~2002), plan to use
their time-domain information to separate quasars from stars. In
addition to color selection, it is expected that the lack of proper
motion and selection by photometric variability properties will result
in fairly clean samples with an efficiency much higher than the 
color-based 73\%.  Also, having an alternative method to color selection is needed to avoid a 
bias against extincted systems with non-standard colors. However, until 
very recently, it was not clear how variability could be used in a quantitative 
fashion to achieve these goals. 

A simple way of using variability to separate quasars from stars is
to take advantage of the fact that quasars are generally more variable 
than stars. Sesar et al.~(2007) demonstrated that at least 90\% of quasars 
have a root-mean-square (rms) variability of at least 0.03
mag. However, the rms by itself does not enable the separation of
quasars from other variable objects such as RR~Lyrae. A recent study
by Schmidt et al.~(2010, hereafter Schm10) showed that by measuring
the rms as a function of time lag ($\Delta t$) between observations,
i.e., the structure function $SF(\Delta t)=A(\Delta t/1 \hbox{yr})^{\gamma}$, 
the parameters $A$ and $\gamma$ enable the separation of quasars and 
stars with great efficiency ($\approx $95\% for ultraviolet-excess 
objects).  This separation is efficient because the rms of quasar 
variability is a few times less on monthly than on yearly time 
scales.  Boutsia et al.~(2009) showed that 
the variability of AGN enables their selection even 
for extended, host-dominated AGN (and X-ray quiet AGN). 

An important piece of information missing from these analyses is the
characteristic time scale for quasar variability, $\tau$.  Studies by
de Vries et al.\ (2005) and Sesar et al.\ (2006) using SDSS combined
with earlier Palomar Observatory Sky Survey measurements for 40,000
SDSS quasars constrained quasar continuum variability on time scales
of 10 to 50 years in the observer's frame. They report that the
characteristic time scale, which in this context is the time lag above
which $SF(\Delta t)$ flattens to a constant value, is of order 1 year
in the quasar rest frame. Recently, Kelly et al.~(2009, hereafter
KBS09) proposed a model where the optical variability of a given
quasar is described by a damped random walk (DRW). The difference with
respect to the well-known random walk is that an additional
self-correcting term pushes any deviations back towards the mean
flux on a time scale $\tau$. It has been established by KBS09,
Koz{\l}owski et al.~(2010, hereafter Koz{\l}10), and MacLeod et
al.~(2010) that a DRW can statistically explain the observed light
curves of quasars at an impressive fidelity level (0.01-0.02 mag). The
model has only three free parameters: the mean value of the light
curve, the driving amplitude of the stochastic process, and the
damping (or characteristic) time scale $\tau$. The predictions are only
statistical, and the random nature reflects our uncertainty about the
details of the physical processes. Koz{\l}10 showed that the DRW model
is an efficient quantitative method to separate quasars from most
variable stars even in the dense stellar environments of the
Magellanic Clouds.    

Here, we test quasar selection in extragalactic fields by applying
the DRW model to the individual light curves of $\sim$10,000 variable point 
sources with $i<19$ from SDSS Stripe 82 (S82). 
This was also recently done by Butler \& Bloom (2010, hereafter B\&B)
using a different approach based on the DRW model. 
About 15\% of these sources are
spectroscopically confirmed quasars whose variability properties have been 
analyzed in detail by MacLeod et al.~(2010). We extend the
selection technique in Schm10 by including information
on the characteristic (damping) time scale, extend the study of 
Koz{\l}10 to an extragalactic survey with extensive spectroscopic
data, and extend the study in B\&B by applying the DRW model to individual 
light curves rather than to the ensemble variability of quasars. 
We describe our variability model in Section~\ref{sec:method}, 
describe the S82 data set in Section~\ref{sec:data}, 
and present our results in terms of efficiency and completeness in Section~\ref{sec:results}.   
In Section~\ref{sec:lsst}, we explore the effects that the light curve length,
cadence, and photometric errors have on the best-fit DRW parameters.  We investigate 
the performance of the quasar selection technique presented here, 
in Koz{\l}10, and in B\&B, for typical (simulated) LSST and Pan-STARRS
cadences. We summarize our findings in Section~\ref{sec:summary}.

\section{Quasar Variability as a Damped Random Walk}
\label{sec:method}
We model the time variability of quasars as a stochastic process 
described by the exponential covariance matrix
\begin{equation}
        S_{ij} = \sigma^2 \exp(-|t_i-t_j|/\tau)
 \label{eqn:cfunc}
\end{equation}
between times $t_i$ and $t_j$.  As detailed by KBS09 and Koz{\l}10,
this corresponds to a DRW (more specifically, an
Ornstein-Uhlenbeck process) with a damping, or characteristic, time
scale $\tau$, and a long-term standard deviation of variability
$\sigma$.   The driving amplitude of short-term variations is defined
as $\hat{\sigma} = \sigma\sqrt{2/\tau}$. Following Koz{\l}10, we model the
light curves and estimate the parameters and their uncertainties using
the method of Press et al.~(1992), its generalization 
in Rybicki \& Press (1992), and the fast computational
implementation described in Rybicki \& Press (1994).  Koz{\l}10 
demonstrate that this approach is more statistically powerful than
the forecasting methods used by KBS09, while still
having computation times scaling linearly with the number of data
points. 

As in MacLeod et al.~(2010), we express the long-term variability in
terms of the structure function (SF), where the SF is the rms 
magnitude difference as a function of the time lag ($\Delta t$)
between measurements. The characteristic time scale for the SF to
reach an asymptotic value $\rm{SF}_{\infty}$ is the damping time scale,
$\tau$. The SF for a DRW is   
\begin{equation}
  SF(\Delta t) = \rm{SF}_{\infty}(1-e^{-|\Delta t|/\tau})^{1/2}, 
\label{eq:sfdt}
\end{equation}
and the asymptotic value at large $\Delta t$ is 
\begin{eqnarray} 
  SF(\Delta t >> \tau) \equiv \rm{SF}_{\infty} = 
  \hat{\sigma}\sqrt{\tau}. 
\label{eq:sfinf}
\end{eqnarray}
For short time lags, 
\begin{eqnarray} 
  SF(\Delta t << \tau) = 
  \rm{SF}_{\infty}\sqrt{\frac{|\Delta t|}{\tau}} = \hat{\sigma}\sqrt{|\Delta t|}. 
\label{eq:sfsmall}
\end{eqnarray}
Therefore, $\hat{\sigma}$ regulates the rise of $SF(\Delta
t<<\tau)$.  The $\hat{\sigma}$ parameter is related to $\gamma$ in the 
equation $SF(\Delta t)=A(\Delta t/1 \hbox{yr})^{\gamma}$, where $A$ and $\gamma$ 
were used to select quasars in Schm10. 
However, $A$ and $\gamma$ do not provide unique information on the 
characteristic time scale $\tau$ (although since 
$\hat{\sigma} \propto \tau^{-1/2}$, there is some information on $\tau$). 
While we obtain individual estimates of $\tau$ and $\hat{\sigma}$ for 
every light curve, in B\&B $\tau$ and $\hat{\sigma}$ are estimated
based on the apparent magnitude, which follows from modeling the
ensemble SF as a DRW as a function of magnitude. We discuss this
difference further in Section~\ref{sec:compareBB}. 

In Section~\ref{sec:results}, we demonstrate that the inclusion of 
$\tau$, in addition to $\hat{\sigma}$, in the selection of quasars 
enables the selection of highly pure samples with high completeness.

\section{           The SDSS Stripe 82 Data Set     }
\label{sec:data}
The Sloan Digital Sky Survey (SDSS, York et al.~2000) provides
homogeneous and deep ($r < 22.5$) photometry in five passbands
($ugriz$, Fukugita et al.~1996; Gunn et al.~1998; Smith et al.~2002).
To test our quasar selection method, we utilize the $g$-band light
curves from the SDSS Stripe 82 (S82), which covers the sky 
region defined by 22hr~24m~$<$~RA~$<$~4hr~8m and 
$-1.27\,^{\circ}<$~Dec~$<1.27\,^{\circ}$ (an area of $\approx$290
deg$^2$). The light curves span about ten years, and the observations are
clustered into yearly seasons about 2--3 months long. There are on
average more than 60 available epochs. Because some observations were
obtained in non-photometric conditions, improved calibration
techniques have been applied to the SDSS S82 data by Ivezi\'{c} et
al.~(2007) and Sesar et al.~(2007), and we use their  results. 
For these data, photometric zero-point errors are 0.01--0.02 mag. 

\begin{figure*}[t!]
\epsscale{.35}
\plotone{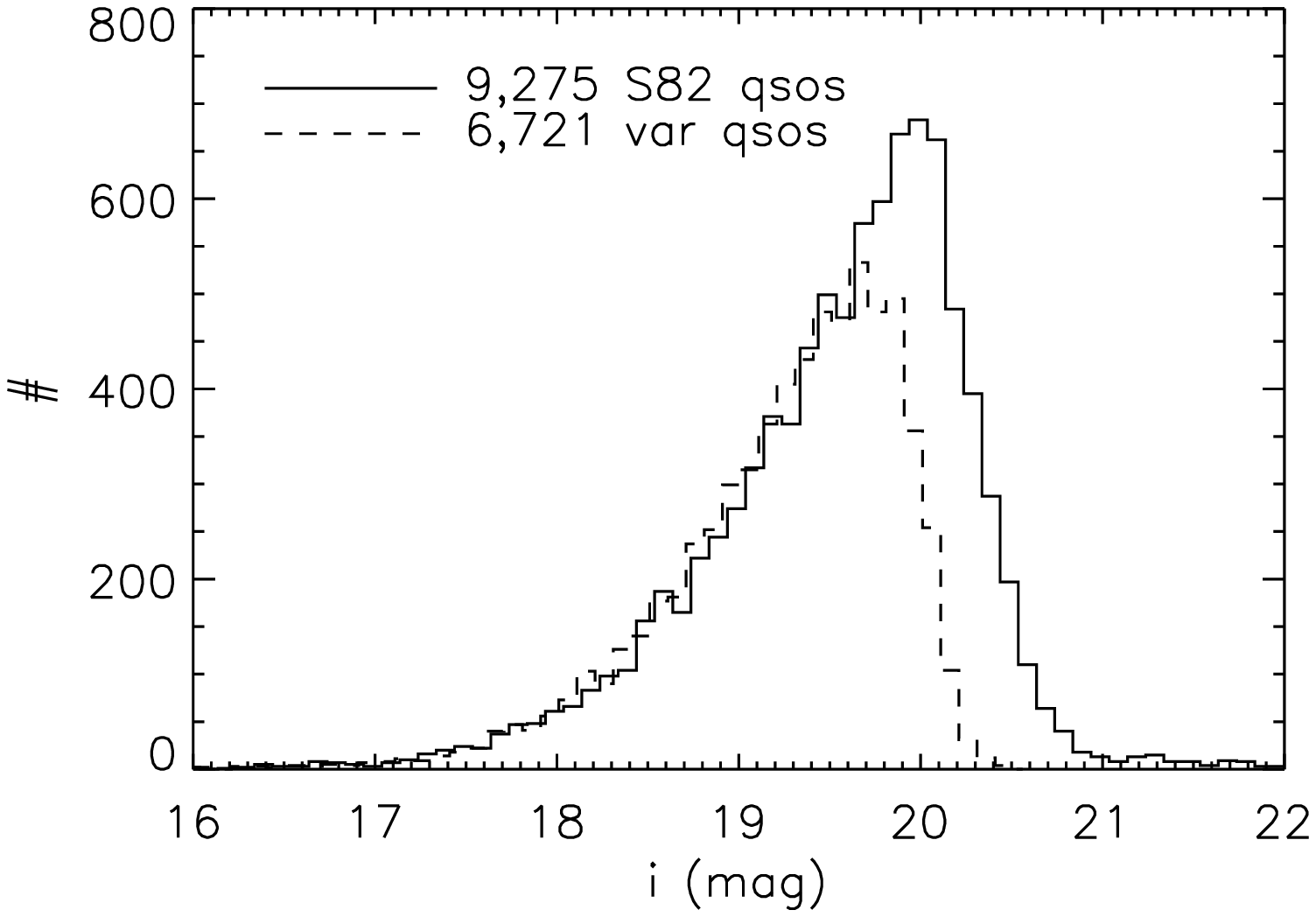}  
\plotone{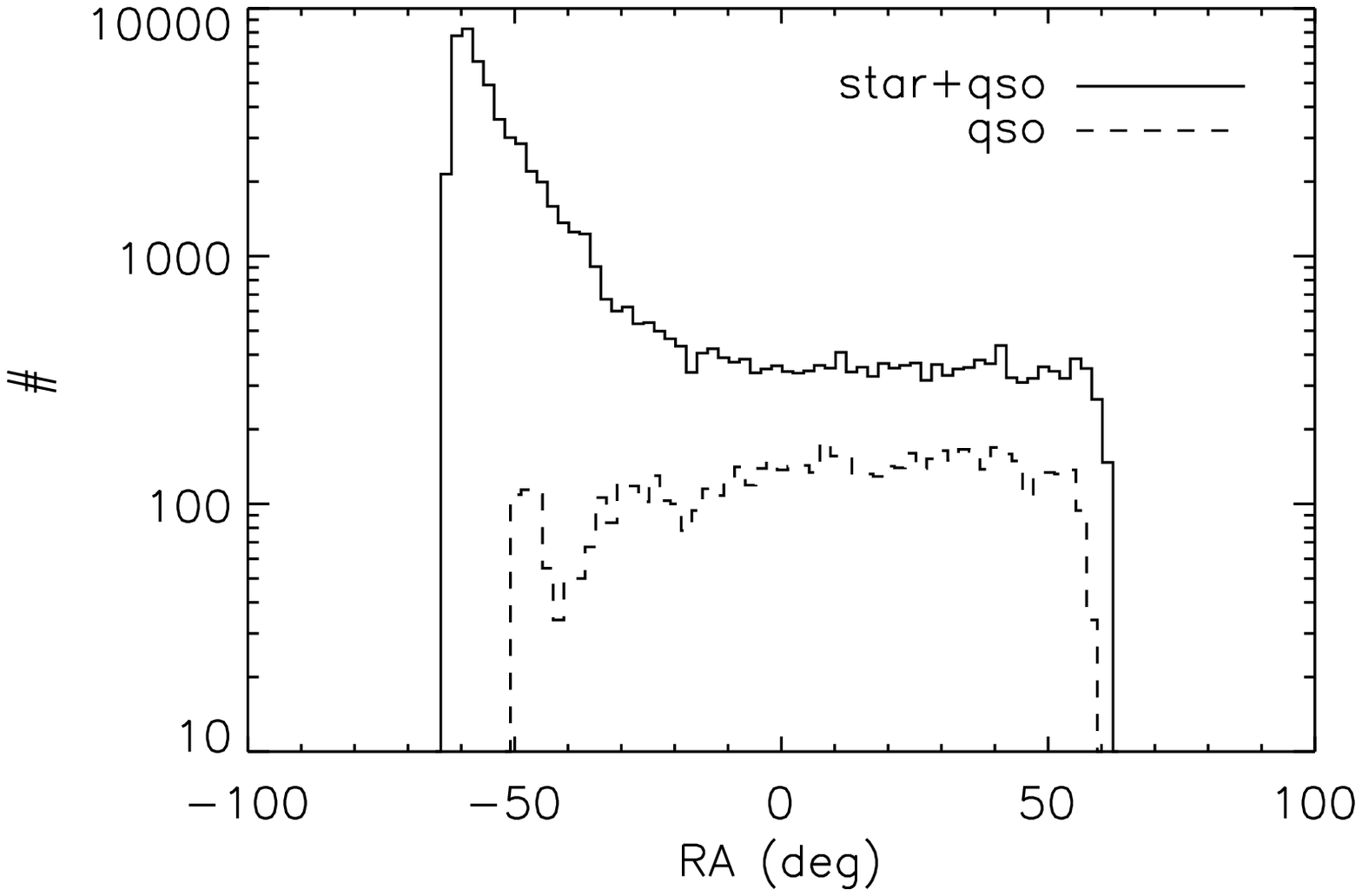}
\caption{\footnotesize \emph{Top panel:} 
  quasar counts as a function of $i$ magnitude for all
  spectroscopically confirmed S82 quasars (MacLeod et al.~2010; solid
  line) and those in the S82 variable point source catalog (Ivezi\'{c}
  et al.~2007; dashed line).   \emph{Bottom panel:} 
  counts for all variable point sources (solid) and for
  spectroscopically confirmed quasars (dashed), as a function
  of Right Ascension. In order to minimize stellar contamination from 
  the Galactic disk, we only include sources with 
  $-35\,^{\circ}<$~RA~$<50\,^{\circ}$ in the subsequent analysis (with 
  Galactic latitudes $-45\,^{\circ}<b<-37\,^{\circ}$).
}
\label{fig:ra_i_hist}
\end{figure*}

We apply the DRW model to all objects in the S82 variable point source
catalog (Ivezi\'{c} et al.~2007). These are objects 
with at least ten observations, an rms of variability in the $g$ 
and $r$ bands exceeding 0.05 mag, and a $\chi^2$ per degree of freedom
exceeding 3 for the light curves in these bands being fit with a
constant flux. We only include observations spaced by at least half a
day apart when applying the model, since we are interested in the
selection of quasars which generally vary on longer time scales than 
variable stars. 
The S82 data set is spectroscopically complete for $i<19$ in the
quasar region of $(u-g)$, $(g-r)$, and $(r-i)$ color space (Richards
et al.~2002; see the top panel of Figure~\ref{fig:ra_i_hist} for the 
total distribution of quasars in $i$). We include newly confirmed DR7
(Abazajian et al.~2009) quasars, and limit our data set to the 52,547
variable objects with $i<19$ when evaluating quasar selection based on
variability (Section~\ref{sec:compeff}).  In this sample, there are
1,912 (4\%) spectroscopically identified quasars.  

The distribution of the known quasars in RA is compared to the
distribution of all variable sources in the bottom panel of
Figure~\ref{fig:ra_i_hist}. The quasar counts fall off at RA~$<-35$
deg, while the full distribution rises due to the rising density of
Galactic stars. Therefore, we further restrict our sample to the
10,024 objects with $-35\,^{\circ}<$~RA~$<50\,^{\circ}$ in order to
minimize the stellar contamination from the Galaxy. 
This sample of 10,024 variable sources, including 1,490 (15\%)
spectroscopically confirmed quasars, defines our final sample when
evaluating quasar selection based on variability in
Section~\ref{sec:compeff}. However, to demonstrate the performance in
regions of increased stellar density, we also compare to the results
for the region $-35\,^{\circ}<$~RA~$<-25\,^{\circ}$. 

\begin{figure*}[t!]
\epsscale{.35}
\plotone{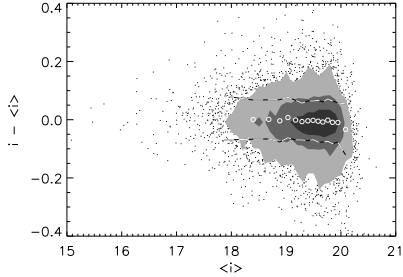}  
\caption{\footnotesize Differences in the SDSS BEST $i$ magnitude and best-fit
  mean $i$ magnitude, $\langle i\rangle$, as a function of 
  $\langle i\rangle$ for S82 quasar light curves modeled as a 
  DRW. The differences arise because the SDSS best $i$ magnitudes are
  for a single epoch. The $\pm 1\sigma$ range is shown with dashed lines.
  The contours show regions containing 25\%, 50\%, and 90\% of the
  total number of data points.}
\label{fig:mui}
\end{figure*}

It is important to note that for variable objects, the single-epoch
$i$ magnitudes can be very different from the best-fit mean magnitudes
from the DRW model, $\langle i\rangle$, since the latter are based on
the entire time series.  In Figure~\ref{fig:mui}, the distribution of
$i-\langle i\rangle$ is shown as a function of $\langle i\rangle$ for
S82 quasars. It is centered on zero with the $\pm 1\sigma$ range
spanning almost $0.1$ mag.  Therefore, these differences are
significant. For subsequent analysis, we use the median PSF $i$-band
magnitudes as listed in the variable point source catalog, which are
similar to the $\langle i\rangle$ values.  All magnitudes are
corrected for interstellar dust extinction using Schlegel et al.~(1998). 

\section{                      Results                               }
\label{sec:results}
\subsection{Variability of Quasars and Other Objects}
In this Section, we present the best-fit variability parameters  
after applying the DRW model to the light curves of all variable point
sources in S82. These parameters are used to develop a method for
separating quasars from other variable point sources. We compare the
results for two subsamples: one containing the spectroscopically
confirmed quasars, and one containing all other objects (which is
dominated by stars). We also investigate in detail the objects in the
latter subsample that have  best-fit time scales greater than 100
days, which is indicative of quasar variability.  

\begin{figure*}[t!]
\epsscale{.8}
\plotone{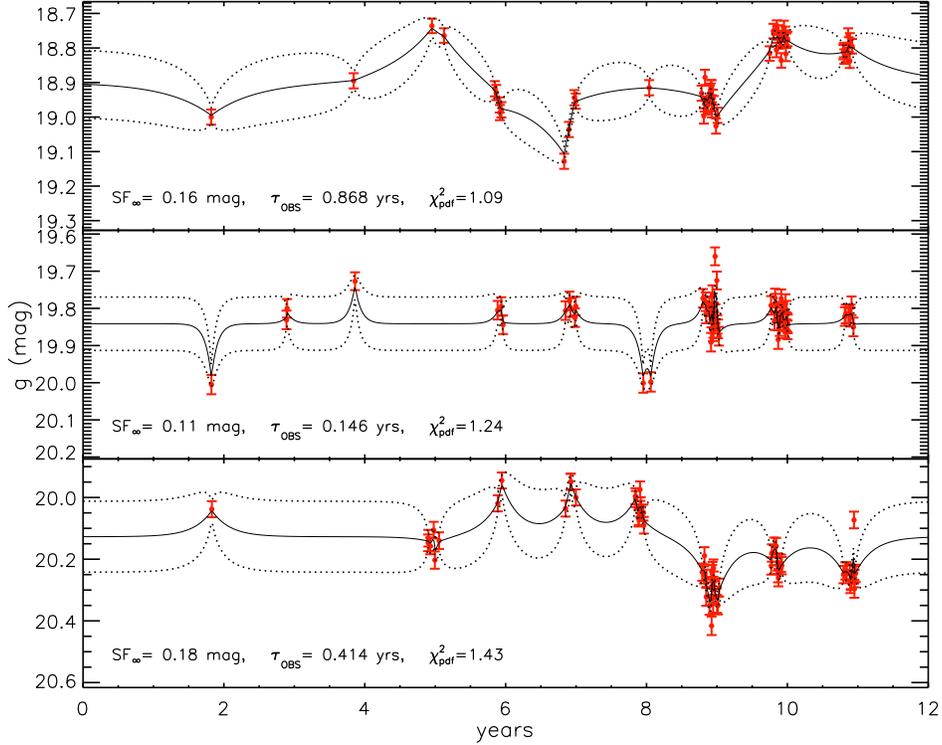}
\caption{\footnotesize The top panel shows a sample light curve for a
  confirmed quasar. Data points with error bars show the	
  observed data, and the solid line shows the weighted average
  of all consistent DRW models. Dotted lines show the $\pm 1\sigma$
  range of these possible stochastic models about the average. The
  middle panel is a similar plot for an unknown object whose light
  curve can only marginally be distinguished from noise 
  ($\Delta L_{\rm{noise}} = 0.4$). The bottom panel is a similar plot for a
  confirmed quasar with an outlying data point around 11~years, leading to a large $\chi^2_{\rm{pdf}}$.  
  The best-fit variability parameters along with $\chi^2_{\rm{pdf}} =
\chi^2/N_{\rm{dof}}$ are listed at the bottom of the panels.}
\label{fig:LCeg}
\end{figure*}

Figure~\ref{fig:LCeg} shows examples of fitted light curves. 
The top panel shows a spectroscopically confirmed
quasar whose variability is well-fit by the model, and the middle panel shows an  
unknown object. The weighted average of all model light curves
consistent with the data is shown, and the ``error snake'' is the 
$\pm 1\sigma$ range of those light curves about this mean. The
best-fit variability parameters along with $\chi^2_{\rm{pdf}} = \chi^2/N_{\rm{dof}}$ are
listed at the bottom of the panels, where $N_{\rm{dof}}$ is the number of
degrees of freedom. Additional information on the goodness of fit is
provided by the parameter  $\Delta L_{\rm{noise}} \equiv \ln{(L_{\rm{best}}/L_{\rm{noise}})}$ 
which was used in Koz{\l}10 and MacLeod et al.~(2010) to select light
curves that are better described by a DRW than by pure white noise. 
Here, $L_{\rm{best}}$ is the likelihood of the best-fit stochastic model and
$L_{\rm{noise}}$ is that for a white noise solution ($\tau \equiv 0$).  
In addition to having a larger $\chi^2_{\rm{pdf}}$, the light curve for the
unclassified object in the middle panel of Figure~\ref{fig:LCeg} has
$\Delta L_{\rm{noise}} = 0.4$ and is therefore only marginally more
consistent with the DRW model than with white noise.  
In some cases (see the bottom panel), outliers in the light curve
cause a relatively large $\chi^2_{\rm{pdf}}$ while the likelihood is
high for a DRW ($\Delta L_{\rm{noise}}=34$). Outliers such as these
could either represent a model failure or a bad data point with a true
error that is larger than the reported value.  A detailed analysis of
these cases will be discussed in a future publication. 

\begin{figure*}[t!]
\epsscale{.7}
\plotone{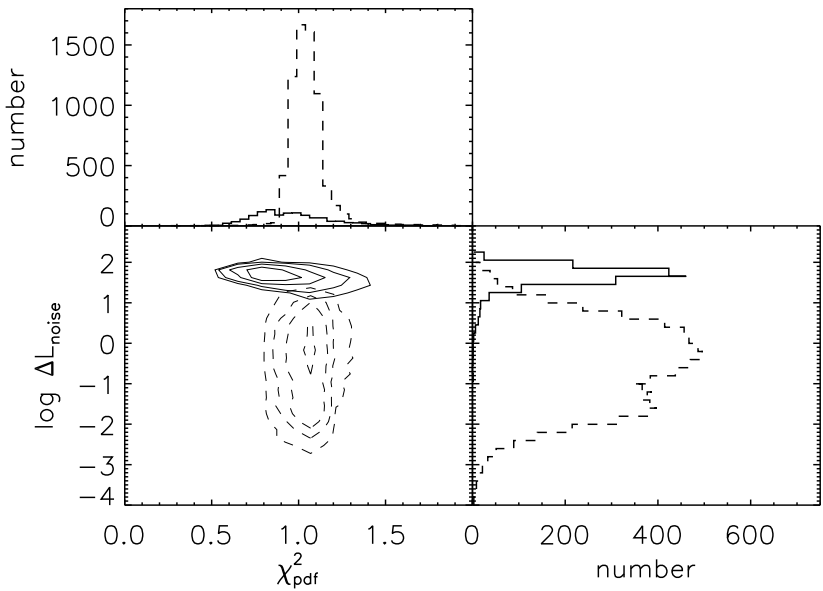}
\caption{\footnotesize  Goodness of fit parameters 
  $\chi^2_{\rm{pdf}}=\chi^2/N_{\rm{dof}}$ versus $\log \Delta L_{\rm{noise}}$ 
  for spectroscopically confirmed quasars (solid) and all other objects
  (dashed) in our S82 data set restricted to $i<19$. The
  contours show regions containing 30, 60, 80, and 90\% of data
  points in each subsample. The marginal distributions are
  shown in the upper and right-hand panels.
}
\label{fig:hist}
\end{figure*}

\begin{figure*}[t!]
\epsscale{.5}
\plotone{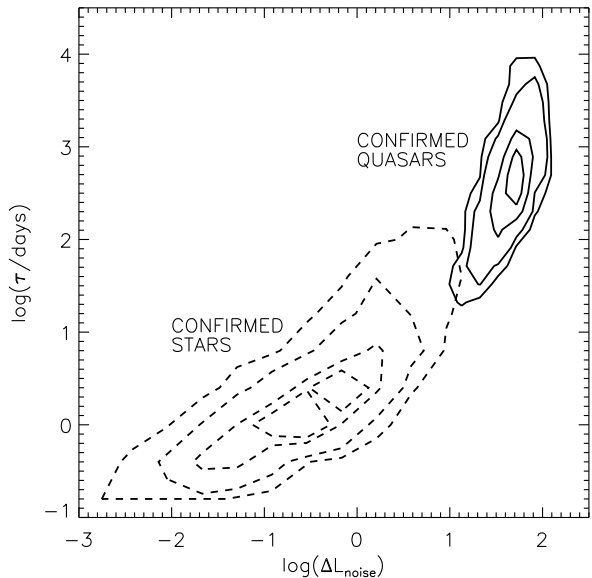}
\caption{\footnotesize Dependence of the best-fit characteristic time
  scale on the relative likelihood $\Delta L_{\rm{noise}}$. Solid (dashed)
  contours show the 25\%, 50\%, 75\%, and 90\% levels for spectroscopically
  confirmed quasars (stars). 
}
\label{fig:tauLnoise}
\end{figure*}

In Figure~\ref{fig:hist}, $\chi^2_{\rm{pdf}}$ is shown as a function of $\Delta
L_{\rm{noise}}$ for the known quasars and all other
objects in the variable S82 sample.
The quasars typically have $\chi_{\rm{pdf}}^2 \sim
0.85$ while all other variable S82 objects typically have
$\chi_{\rm{pdf}}^2 \sim 1.0$, with a large overlap
between the two distributions. The overlap is greatly reduced when
considering the marginal distributions of $\Delta L_{\rm{noise}}$.  
While $\chi^2_{\rm{pdf}}$ barely resolves two populations (quasars and
stars), $\Delta L_{\rm{noise}}$ efficiently separates quasars
(which are well-fit by the stochastic model) 
from variable stars (which are better fit by pure noise).  Objects with high 
$\Delta L_{\rm{noise}}$ also tend to vary on longer time scales (see
Figure~\ref{fig:tauLnoise}). Therefore, the characteristic time scale
$\tau$ also provides an efficient way to distinguish quasars 
from stars. 

\begin{figure*}[t!]
\epsscale{.7}
\plotone{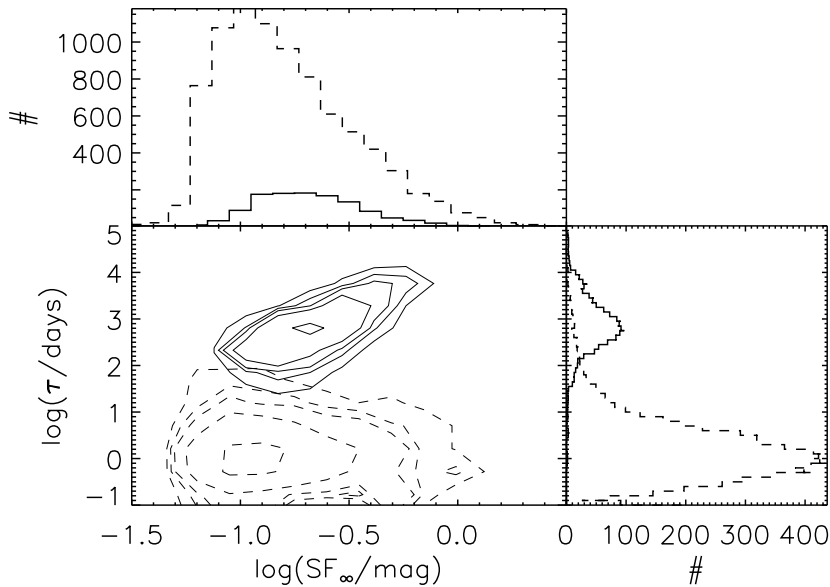}
\caption{\footnotesize Characteristic time scale $\tau$ versus
  asymptotic rms variability $\rm{SF}_\infty$ for spectroscopically
  confirmed quasars (solid) and all other objects (dashed) in our
  S82 data set restricted to $i<19$. The contours show regions
  containing 10, 60, 75, 85, and 90\% of data points in each subsample. 
  The marginal distributions are shown in the upper and right-hand panels. }
\label{fig:sftau}
\end{figure*}

The distributions of $\tau$ and $\rm{SF}_{\infty}$ for the quasar and  
stellar subsamples are shown in Figure~\ref{fig:sftau}. The $\tau$  
distribution (in the observer's frame) peaks near 500 days for
the known quasars and around 1 day for the other objects,
providing an efficient and simple way to distinguish between the two
subsamples. 
We note that because quasars are known to have aperiodic variability, 
eliminating periodic sources should lead to a cleaner sample of quasars.  
We could classify variable sources as periodic if the
likelihood for the peak in the Lomb--Scargle periodogram (as
implemented by Press et al.~1992b) being observed at random
is $\log{(p_{\rm{periodic}})}<-3$ and the estimated period is less
than $\sim$200~days. For longer periods, many quasars are falsely
flagged as periodic because the periodogram only tests for whether a
sine wave is a better fit than a constant, and this is frequently true
for quasars with only a few ``oscillations'' in the data (see Koz{\l}10).
Because we have few longer period variable stars in S82, adding such a
cut for periodic stars adds little and so we have not included
it. This differs from the Magellanic Cloud fields examined by
Koz{\l}10 where there are many longer-period periodic stars (e.g.,
Cepheids). 

\begin{figure*}[h!]
\centerline{
\includegraphics[width=2.5in]{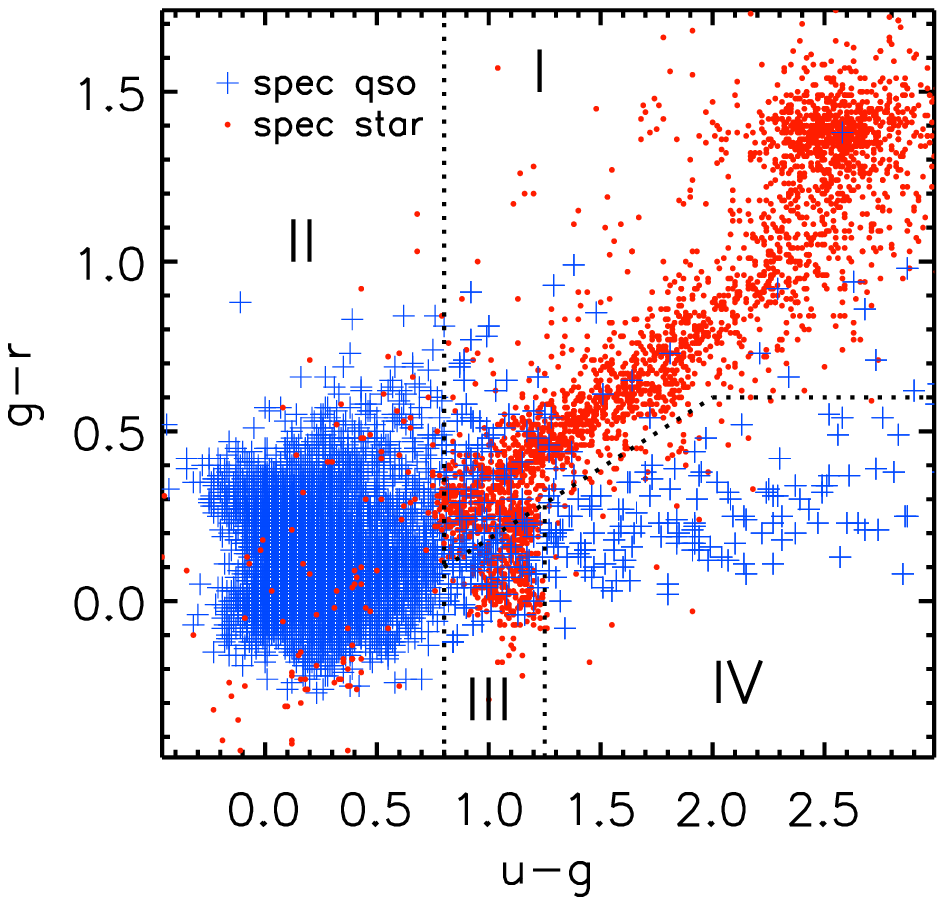}
\includegraphics[width=2.5in]{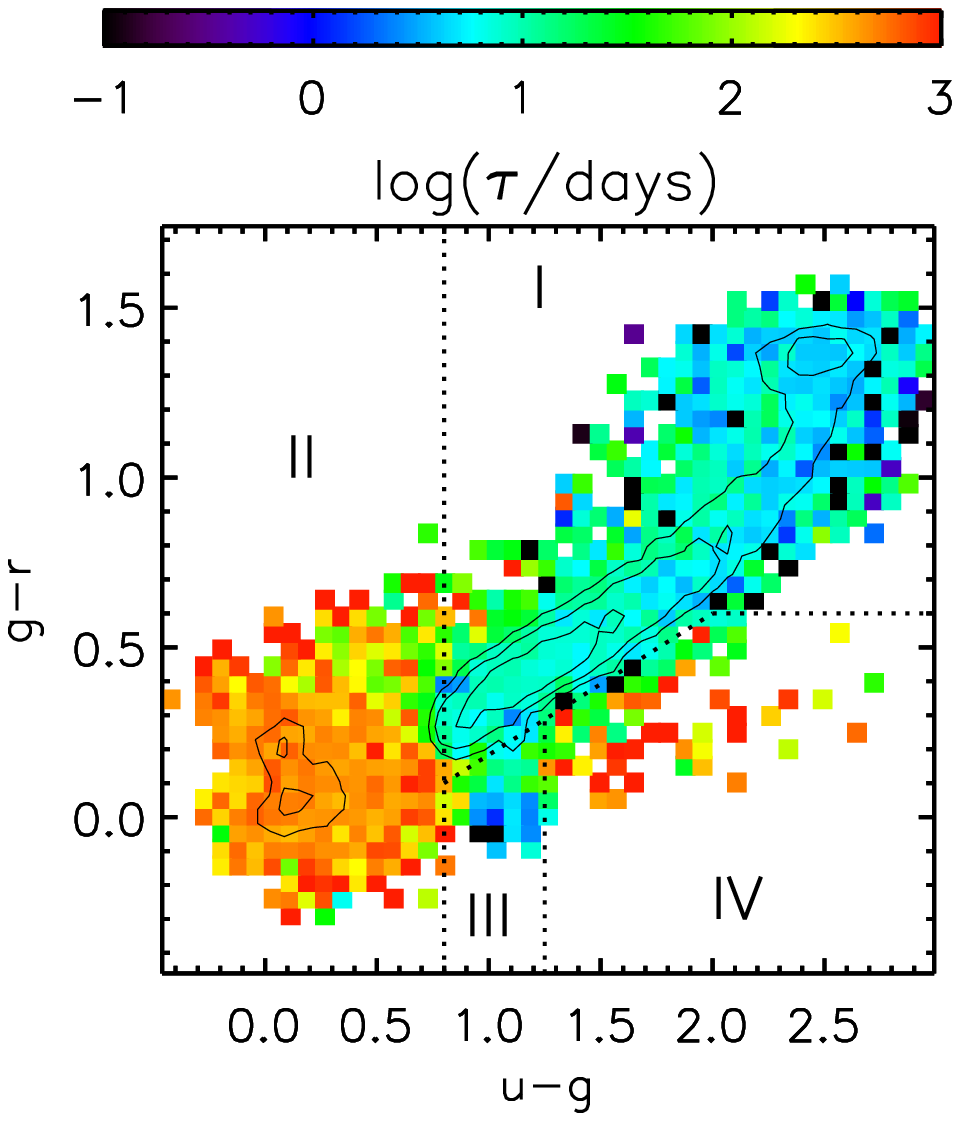}}\centerline{
\includegraphics[width=2.5in]{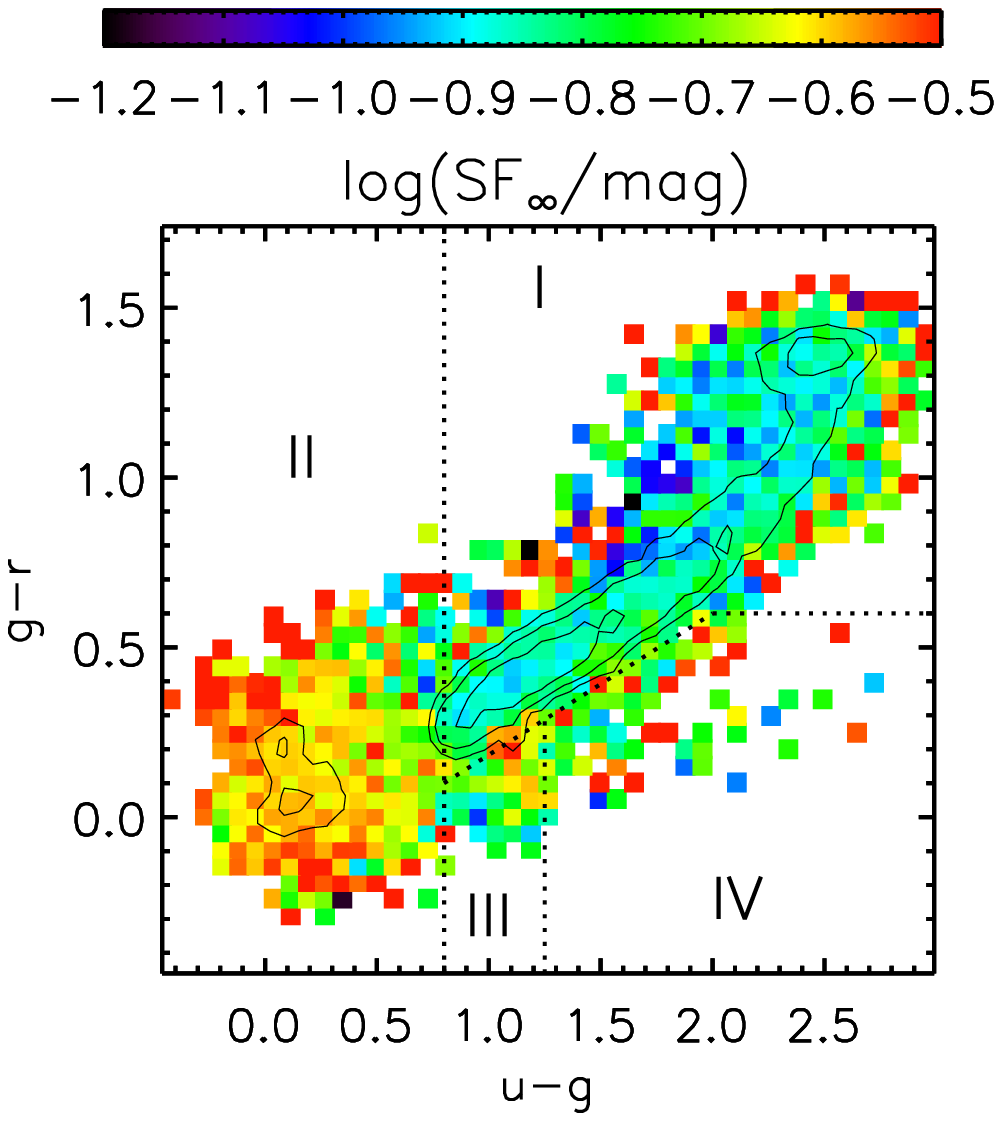}
\includegraphics[width=2.5in]{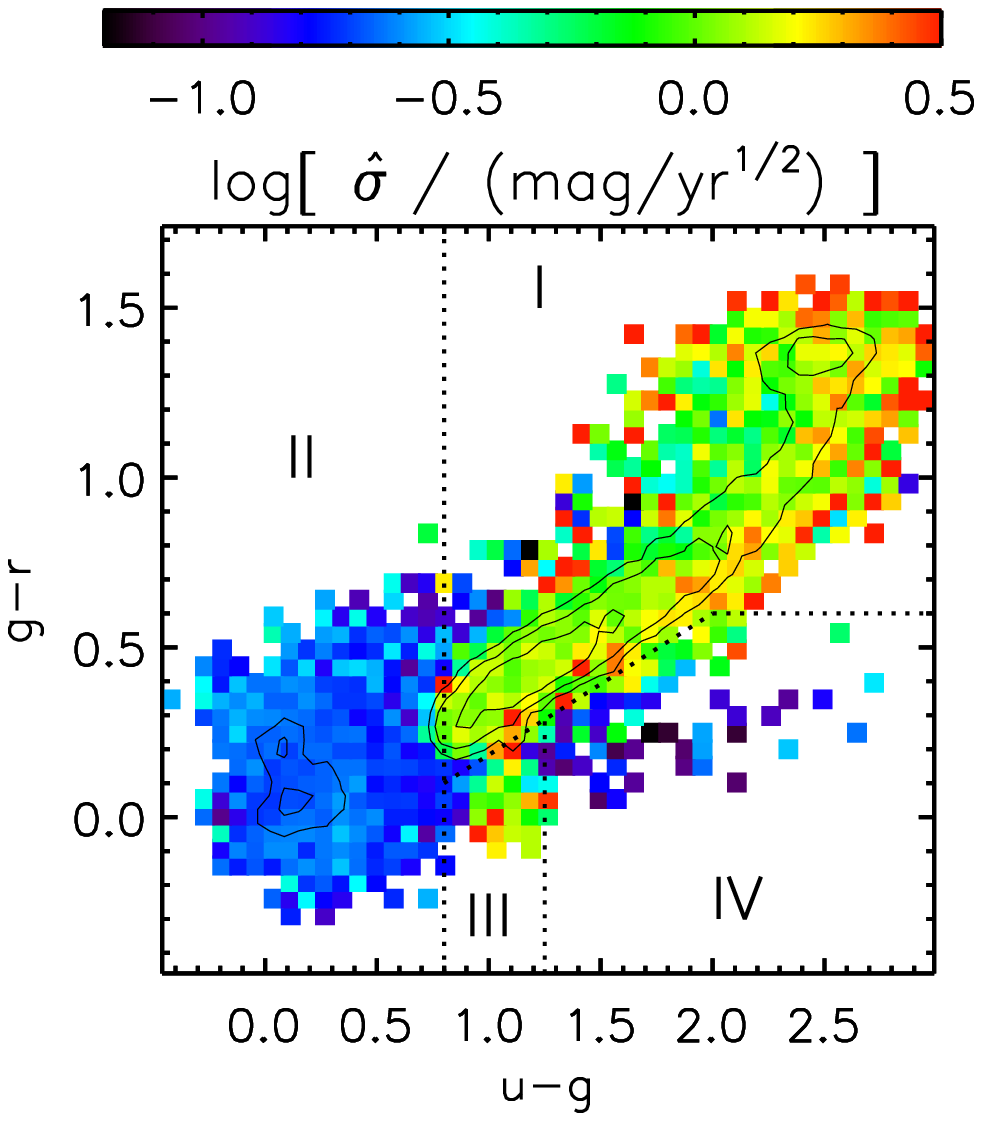}}
\caption{\footnotesize \emph{Top-left:} 
  $u-g$ versus $g-r$ colors of all spectroscopically confirmed
  S82 quasars (crosses) and spectroscopically confirmed stars
  (dots). 
  \emph{Top-right:}  The median characteristic time
  scale $\tau$ for all variable point sources in S82 that are well-described
  by the DRW model ($\Delta L_{\rm{noise}}>2$), as a
  function of $u-g$ and $g-r$, following the legend at the
  top. The contours show regions containing 
  25\%, 50\%, and 75\% of the total number of data points. 
  The red regions with large $\tau$ coincide with the quasar
  color regions, providing a convenient method for quasar
  selection based on variability. 
  \emph{Bottom-left:} As in top-left panel but for
  the asymptotic rms of variability, $\rm{SF}_{\infty}$. 
  \emph{Bottom-right:} As in top-left panel but for the driving
  amplitude of short-term variations, $\hat{\sigma}=\rm{SF}_{\infty}/\sqrt{\tau}$.
  The dotted lines divide the color space into four
  characteristic regions: the stellar locus (I), the low-redshift
  ($z\lesssim 2$) quasar region (II), the RR~Lyrae region (III), and
  the high-redshift ($z\gtrsim 3.2$) quasar region (IV). 
}
\label{fig:colormap}
\end{figure*}

The use of $\tau$ as an efficient classifier is further illustrated in 
Figure~\ref{fig:colormap}. 
The top-left panel shows the $(u-g)$ and $(g-r)$ colors for
spectroscopically confirmed quasars and stars. 
The top-right panel shows the median value of $\tau$ as a function of
$(u-g)$ and $(g-r)$ color. This panel shows that the quasar locus
(region II) as well as the high redshift quasars ($z\gtrsim 3.2$; region IV) are 
dominated by long time scales, while the stellar locus (region I) and 
the RR~Lyrae (region III) have short time scales. The time scale
information is able to clearly distinguish between quasars and RR
Lyrae, as opposed to when using the rms of variability alone (see
Figure 3 in Sesar et al.~2007).   
The quasar regions of color space also tend to have larger
median values of $\rm{SF}_{\infty}$, as seen in the bottom-left panel 
of Figure~\ref{fig:colormap}. The remaining panel shows the median
driving amplitude of short-term variations,
$\hat{\sigma}=\rm{SF}_{\infty}/\sqrt{\tau}$, against $(u-g)$ and $(g-r)$
color. Because it is anti-correlated with $\tau$ (see Koz{\l}10), the
quasar regions show systematically smaller $\hat{\sigma}$ than the
stellar regions.  

\begin{figure*}[t!]
\epsscale{.6}
\plotone{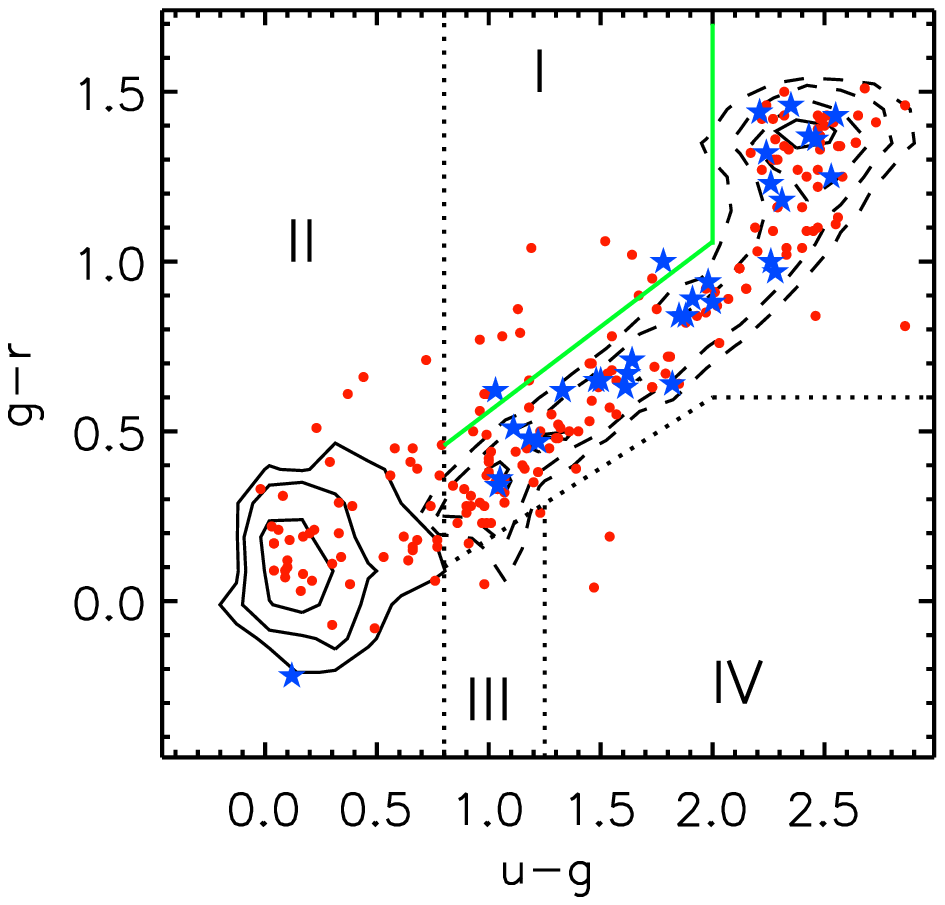}
\caption{\footnotesize 
  $u-g$ versus $g-r$ colors of bright ($i<19$) variable point sources
  with $\tau\geq 100$~days (solid contours) and $\tau<100$~days
  (dashed; 25\%, 60\%, and 75\% levels). Also shown are the
  233 objects with $100 \leq \tau < 10^5$~days and $\Delta L_{\rm{noise}}>2$ 
  that are not spectroscopically confirmed quasars (red dots). Those
  with confirmed stellar origin are represented by blue stars. We use
  the solid line in region I to define a subsample of objects that lie
  off the stellar locus.} 
\label{fig:uggrtimes}
\end{figure*}

\begin{figure*}[h!]
\centering
\includegraphics[width=3.5in]{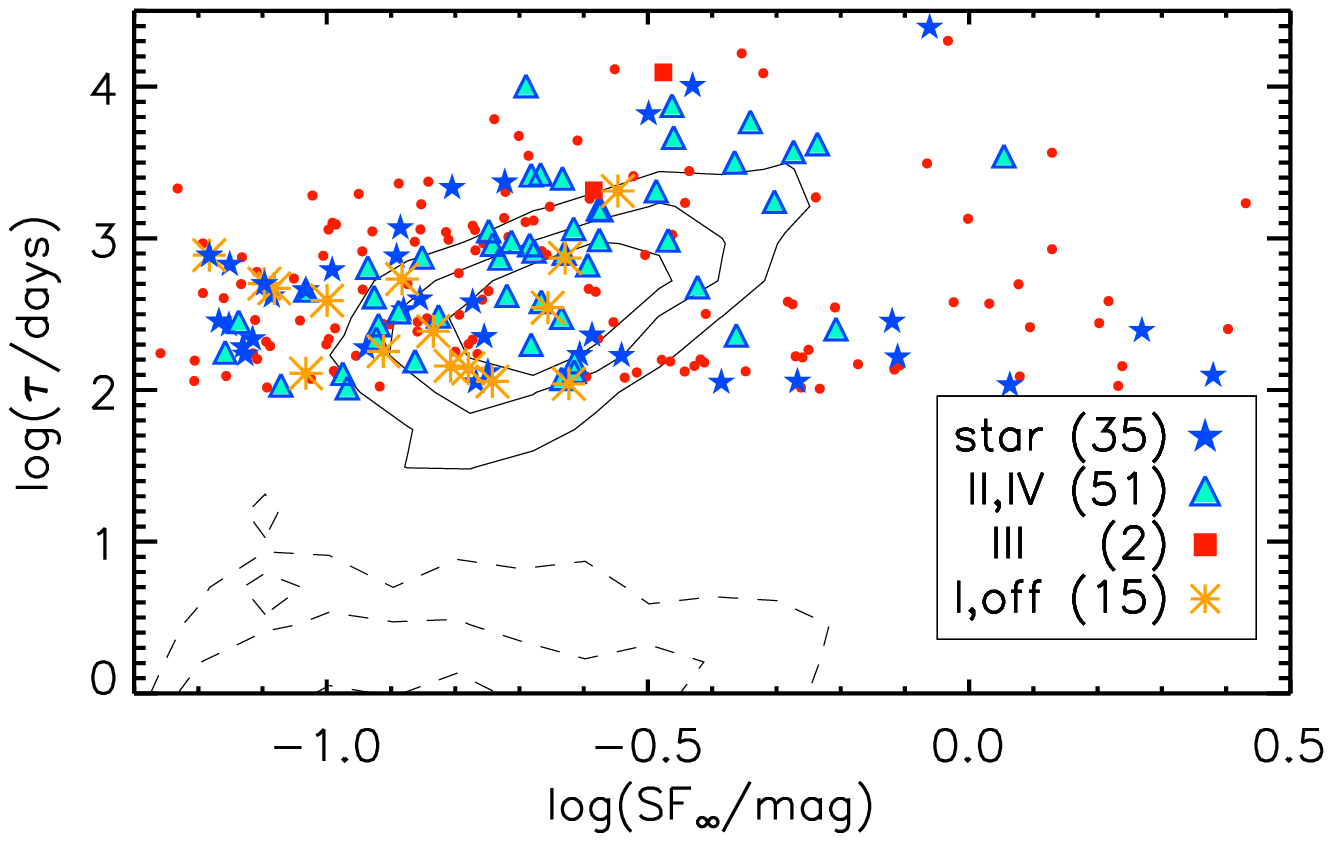}
\includegraphics[width=3.5in]{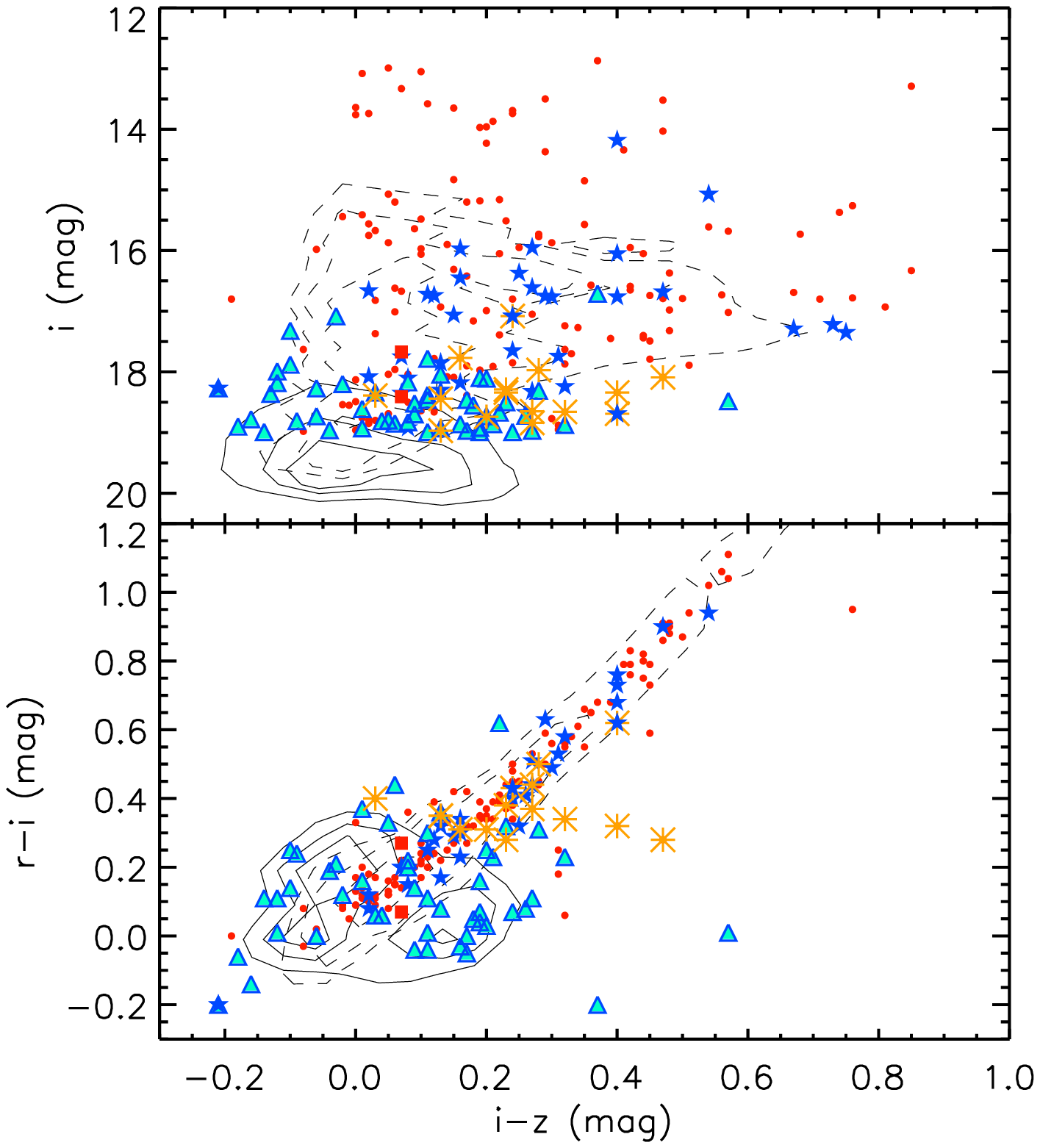}
\caption{\footnotesize \emph{Top}:  Variability time scale $\tau$
  versus amplitude $\rm{SF}_{\infty}$ for all spectroscopically confirmed
  quasars (solid contours), all confirmed stars (dashed), and 
  the 233 objects with $100 \leq \tau < 10^5$~days, 
  $\Delta L_{\rm{noise}}>2$, and $i<19$ that are not already confirmed quasars (symbols). 
  Blue stars represent those with confirmed stellar origin,
  and cyan triangles represent those with quasar-like colors (regions II or IV
  of Figure~\ref{fig:colormap}). Red squares represent those in
  region III, and orange asterisks represent those that lie in region I but above the solid line in
  Figure~\ref{fig:uggrtimes}. The number of objects in each subset is
  listed in the legend, and the remaining 130 objects are represented
  by red dots. The contour levels are 25\%, 50\%, and 75\%.
  \emph{Middle and bottom}: As in top panel but for $i$ magnitude and
  $r-i$ color, respectively, as a function of $i-z$. 
}
\label{fig:tauhigh}
\end{figure*}

Figure~\ref{fig:uggrtimes} compares the distribution of objects with
$\tau\geq 100$~days and $\tau<100$~days in $(u-g,g-r)$ space, as well
as the subset of 233 objects with  $100 \leq \tau < 10^5$~days and 
$\Delta L_{\rm{noise}}>2$ that are not spectroscopically confirmed
quasars. The latter contains good quasar candidates, especially those
in regions II or IV (51 objects). Thirty-five of these are
spectroscopically confirmed stars, represented by the star symbols. 
Figure~\ref{fig:tauhigh} shows the distribution of these 233 objects in 
$(\tau,\rm{SF}_{\infty})$, $(i,i-z)$, and $(r-i,i-z)$ space. The
subset in regions II or IV, represented by triangles, are clustered
near the quasar region at faint $i$ and blue $(r-i,i-z)$ colors. We
inspected their observed and phased light curves (using the
Supersmoother algorithm, see Reimann~1994), and they appear consistent  
with quasar light curves with no signs of periodicity. 
About half of these 51 objects are detected in 
GALEX\footnote{\footnotesize Galaxy Evolution
  Explorer; see http://galex.stsci.edu/GR6/. A matching radius of
  $3''$ was used.}  GR6 with GALEX-SDSS UV-optical colors
resembling those of quasars 
($n-u < 1.5$ and $u-g < 0.8$, or $f-n > 0$ and $g-r < 0.5$; see
{Ag{\"u}eros} et al.~2005). About 15\% have
2MASS\footnote{\footnotesize Two Micron All Sky Survey; see
  http://www.ipac.caltech.edu/2mass/. A matching radius of $1.5''$
was used.}  colors $J-K>1$ that are also indicative quasar nature.  
We also
investigated the 15 unknown objects with $\tau\geq 100$~days 
that lie in region I but off the stellar locus (above the
solid line in Figure~\ref{fig:uggrtimes}).  These objects (represented by 
asterisks in Figure~\ref{fig:tauhigh}) seem consistent with
reddened quasars, as their light curves are consistent with quasars,
and they are rather faint and clustered at redder $(r-i,i-z)$
colors. However, their median $\rm{SF}_{\infty}$ (0.15 mag) is lower
than that for the quasar candidates represented by triangles
(0.22~mag). None of these objects have GALEX-SDSS UV-optical colors
indicative of quasars, but about 30\% have $J-K>1$. Follow-up
spectroscopy will be useful for classifying these two subsamples.  
 
\begin{figure*}[h!]
\epsscale{.6}
\plotone{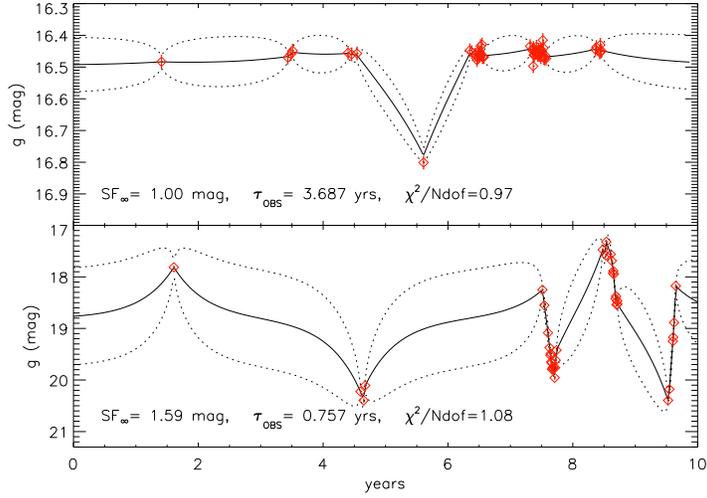}
\caption{\footnotesize 
  Light curves for SDSS J001014.28+005938.6, a probable example of
  a false variable created by a bad data point (top), and SDSS
  J220514.58+000845.7, a probable example of a non-periodic variable
  star (bottom).  Data points with error bars show the	 
  observed data, and the solid line shows the weighted average
  of all consistent DRW models. Dotted lines show the $\pm 1\sigma$
  range of these possible stochastic models about the average. 
  The best-fit variability parameters are listed at the bottom of the panels.
}
\label{fig:eb_agb}
\end{figure*}

The remaining objects with $\tau\geq 100$~days lying on the stellar
locus or in region III seem consistent with stars based on their
$i$ magnitude and colors, and many are spectroscopically confirmed as 
stars. The majority have small amplitudes ($\rm{SF}_{\infty}\approx 0.1$ 
mag). However, there is a subset with large amplitudes
($\rm{SF}_{\infty}>0.8$ mag). The majority of these sources have
a single outlier at a significantly fainter magnitude in their light
curves (e.g., SDSS J001014.28+005938.6, see Figure~\ref{fig:eb_agb})
causing the model to fit a long time scale.  These could be wide
eclipsing binaries or simply a bad data point.  If we reject the 
data point with the largest magnitude difference from the median 
in each light curve with $\rm{SF}_{\infty}>0.8$ mag, and then refit the
models, half now have $\rm{SF}_{\infty}<0.2$ mag and 60\% now have $\Delta L_{\rm{noise}} < 2$
and $\tau < 100$~days. There is a subset of these sources with large
amplitude semi-periodic outbursts, suggestive of AGB stars (e.g.,
SDSS J220514.58+000845.7, see Figure~\ref{fig:eb_agb}). 
In general, however, inspection of both the phased and observed light
curves revealed no examples of truly periodic behavior. A few (2\%) of
these objects have GALEX-SDSS UV-optical colors indicative of quasars,
and about 5\% have $J-K>1$.  

We compile a list of quasar candidates by removing those that already
have SDSS spectra from the list of 233 and including cases where 
$\tau$ has saturated to $10^5$~days. We retain 255 objects lacking
SDSS spectra that have $i<19$, $\Delta L_{\rm{noise}} > 2$, and 
$\tau \geq 100$~days, leading to a sample of 255.  The coordinates, SDSS photometry, and DRW
parameters for each target are provided in Table~\ref{tab:targets}.
These represent our target sample for spectroscopic follow-up.  Note
that 30\% of the target sample fails the $\Delta L_{\rm{noise}} > 2$
and $\tau \geq 100$~day criteria when omitting the most outlying data
point in the light curve.

\begin{deluxetable}{c c c c c c c c c c c}
\tabletypesize{\footnotesize}
\tablewidth{0pt}
\tablecaption{Targets for Spectroscopic Follow-up \label{tab:targets}}
\tablehead{
ID       & RA (deg)  & Dec (deg)   & $i$ & $u-g$ & $g-r$ & $r-i$ &
$i-z$ & log($\tau/$days) & $\rm{SF}_{\infty}$ (mag) & Outlier 
}
\startdata
7912248 & 340.000458 & $\phantom{-}0.304811$ & 15.73 & 2.32 & 1.43 & 1.24 & 0.68 & 2.34 & 0.26 & 1 \\
46516   & $\phantom{00}$3.922421   & $-0.331890$ & 17.97 & 2.02 & 0.87 & 0.35 & 0.19 & 2.06 & 0.06 & 0 \\
48527   & $\phantom{00}$2.699158   & $-0.234567$ & 13.67 & 1.95 & 0.70 & 0.22 & 0.11 & 5.00 & 1.39 & 1 \\
\tableline
\enddata
\tablecomments{Stripe 82 object ID, J2000 coordinates, SDSS photometry, and DRW parameters for 
  255 unknown objects with $\tau \geq 100$~days, $\Delta L_{\rm{noise}} > 2$,
  and $i<19$. For some objects, the 
  time scale has saturated to $\tau=10^5$~days due to insufficient
  light curve lengths. ``Outlier'' is set to 1 for objects that
  fail the $\tau \geq 100$~day and $\Delta L_{\rm{noise}} > 2$
  selection criteria after rejecting the data point furthest from the median
  magnitude.  (See the online version for the complete table.)}
\end{deluxetable}

\clearpage
\subsection{Completeness and Efficiency}
\label{sec:compeff}
\begin{figure*}[t!]
\epsscale{.6}
\plotone{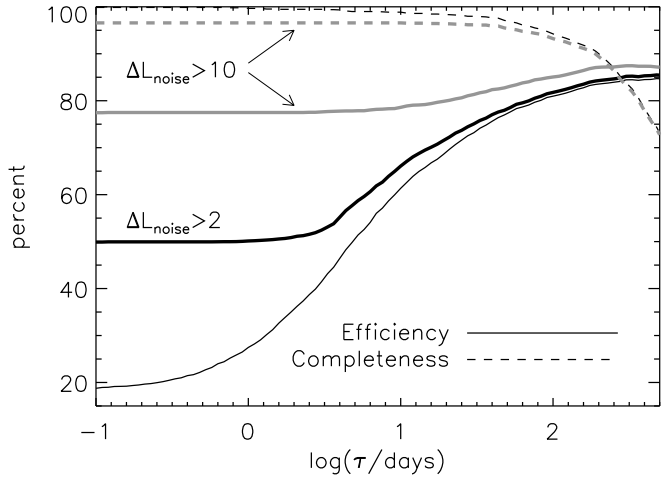}
\caption{\footnotesize $E$ (solid curves) and $C$
  (dashed curves) measured as a function of the minimum threshold
  $\tau$ for variable S82 point sources with $i<19$ and 
  $-35\,^{\circ}<$~RA~$<50\,^{\circ}$.  The thin, solid and
  dashed black curves show the results when using a $\tau$ criterion alone.
  The thick, solid black curve shows the effect of adding a criterion
  of $\Delta L_{\rm{noise}}>2$ ($C$ is unaffected), and the gray
  curves show the effect of adding $\Delta L_{\rm{noise}}>10$.} 
\label{fig:cetau}
\end{figure*}

 To measure the success of our quasar selection, 
 we report the completeness ($C$) and efficiency ($E$), defined by 
\begin{eqnarray}
C = \frac{\#~of~selected~confirmed~quasars}{total~\#~of~confirmed~quasars} \times 100  \nonumber \\
E = \frac{\#~of~selected~confirmed~quasars}{total~\#~of~selected~objects} \times 100,
\end{eqnarray}
for a variety of selection criteria, where ``confirmed'' means that
the source is spectroscopically identified as a quasar. Here, we only 
consider sources with $i<19$, and $E$ is a lower limit for the true
efficiency because the sample of confirmed quasars may not be 100\%
spectroscopically complete. These quantities for a quasar selection
based on $\tau$ alone are shown in Figure~\ref{fig:cetau} as a
function of the minimum threshold $\tau$.  The two quantities are
anti-correlated due to the increased overall fraction of selected
quasars as well as the increased number of missed quasars when
restricting $\tau$ to longer time scales. The completeness drops to
94\% (where $E=81$\%) at a threshold of $\tau>100$~days, beyond which
$C$ dramatically decreases. At the same time, $E$ for $\tau>100$~days
is almost as high as its asymptotic value (85\%). Therefore, we adopt
$\tau\geq 100$~days as the optimal cut for selecting quasars
based on time scale alone.  An additional constraint of $\Delta
L_{\rm{noise}}>2$ (thick line in Figure~\ref{fig:cetau}) does not
affect the resulting $C$ or $E$ by more than 1\% after requiring $\tau
\geq 100$~days, while the constraint $\Delta L_{\rm{noise}}>10$ (gray
lines) increases $E$ with a minor drop in $C$. 

Table~\ref{tab:cuts} lists $C$ and $E$ for various selection criteria,
beginning with the $i$ magnitude and RA limits that define our
starting sample (first row, hence the 100\% completeness), and the 
restrictions $\tau\geq 1$ and 100 days (second and third rows).  
When excluding the objects that fail the $\Delta L_{\rm{noise}} > 2$ and
$\tau \geq 100$~day criteria after omitting the most outlying data
point in the light curve, $E$ rises from 81\% to 87\%, while $C$ only drops
by 1\% (fourth row). This cut is useful for eliminating potentially
bad data in order to boost the efficiency of the selected sample. 
$C$ can be increased to 96\% while maintaining $E=87$\% by reducing 
the limit to $\tau \geq 10^{1.5}$~days and requiring $\Delta L_{\rm{noise}} > 10$ 
(fifth row). Here, we have excluded objects that drop out of the
selection after removing the most outlying data point. When further
restricting the variability-selected sample   
(with $\tau \geq 100$~days) to those objects which have quasar colors
(regions II and IV in Figure~\ref{fig:colormap}), $C$ decreases to
91\% while $E$ rises to 96\% (sixth row; these results are unaffected
when rejecting light curve outliers). 
Table~\ref{tab:cuts} also lists the percentage of each selected sample
that have SDSS spectra, and the breakdown into quasars, stars, and
galaxies based on their SDSS spectroscopic class.  

We also compute $E$ and $C$ for the ultraviolet-excess (``UVX'') and
non-ultraviolet-excess (``nUVX'') color boxes as defined in Schm10.
The former should be spectroscopically complete to $i<19$ (see
Richards et al.~2002).  The latter corresponds to the quasar redshift
regime of $\sim$3, where color selection largely fails.  Here, $C$ is the
percentage of all confirmed variable (n)UVX quasars
that have $\tau \geq 100$~days. For UVX objects, $C=95$\% and  
$E=97$\% for the $\tau \geq 100$~days selection criteria (eighth row of
Table~\ref{tab:cuts}).  For comparison, the Schm10 selection method,
which does not use the $\tau$ information, gives $C=90$\% and 
$E \simeq 95$\%.  For nUVX objects (only 52 objects; tenth row),
$C=100$\% for the $\tau \geq 100$~days selection criteria, and $E$ is
only 69\% due to the fact that only 69\% have spectra.  

\begin{figure*}[h!]
\centerline{
\includegraphics[width=2in]{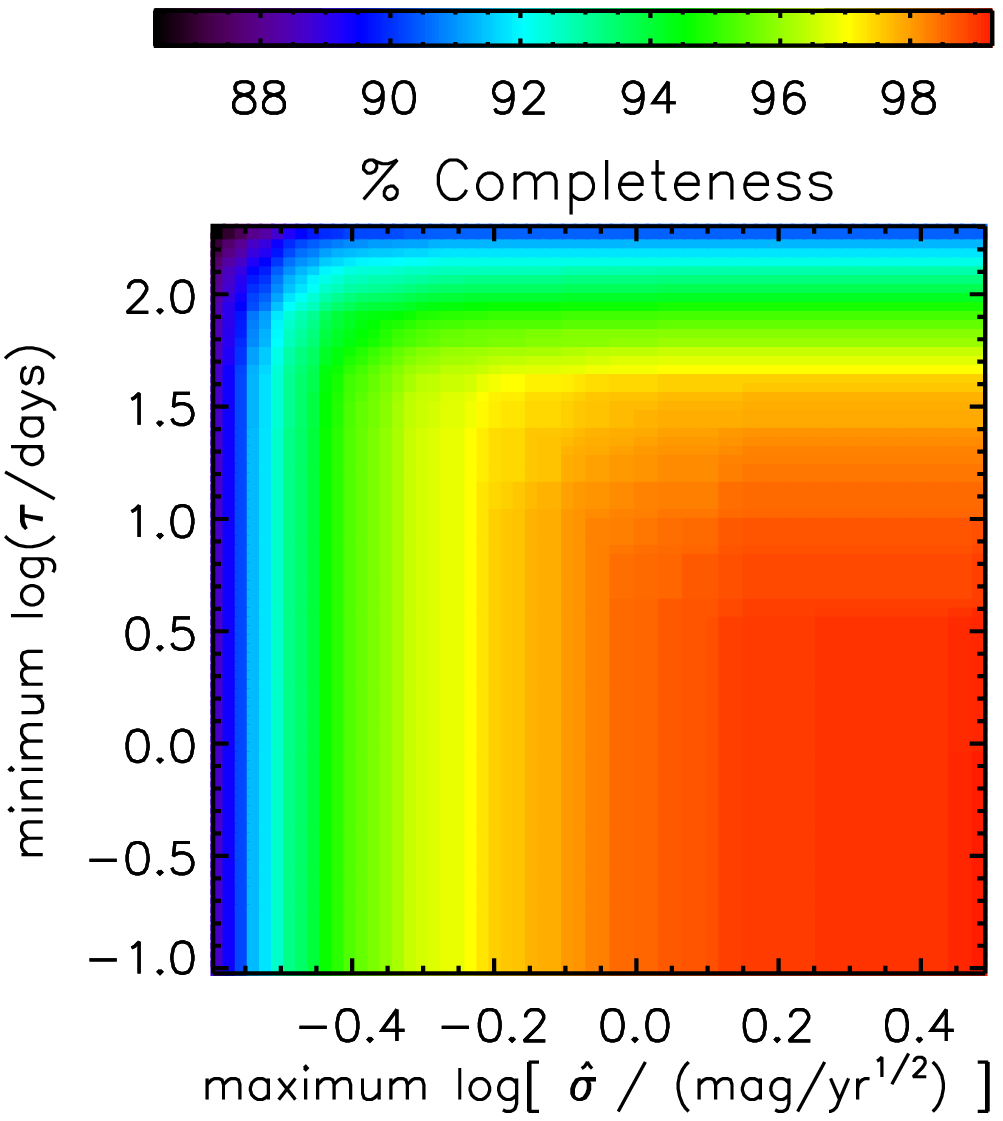}
\includegraphics[width=2in]{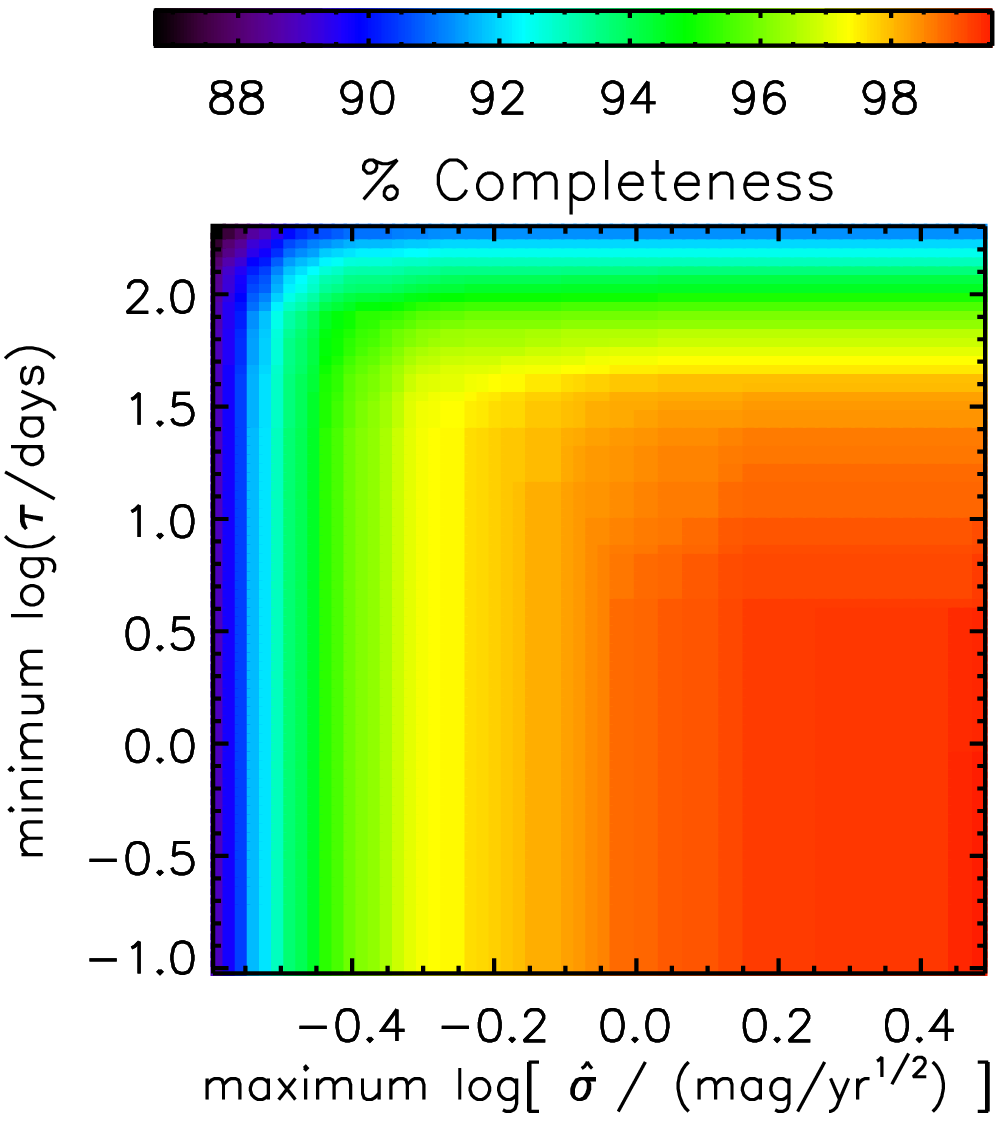}}\centerline{
\includegraphics[width=2in]{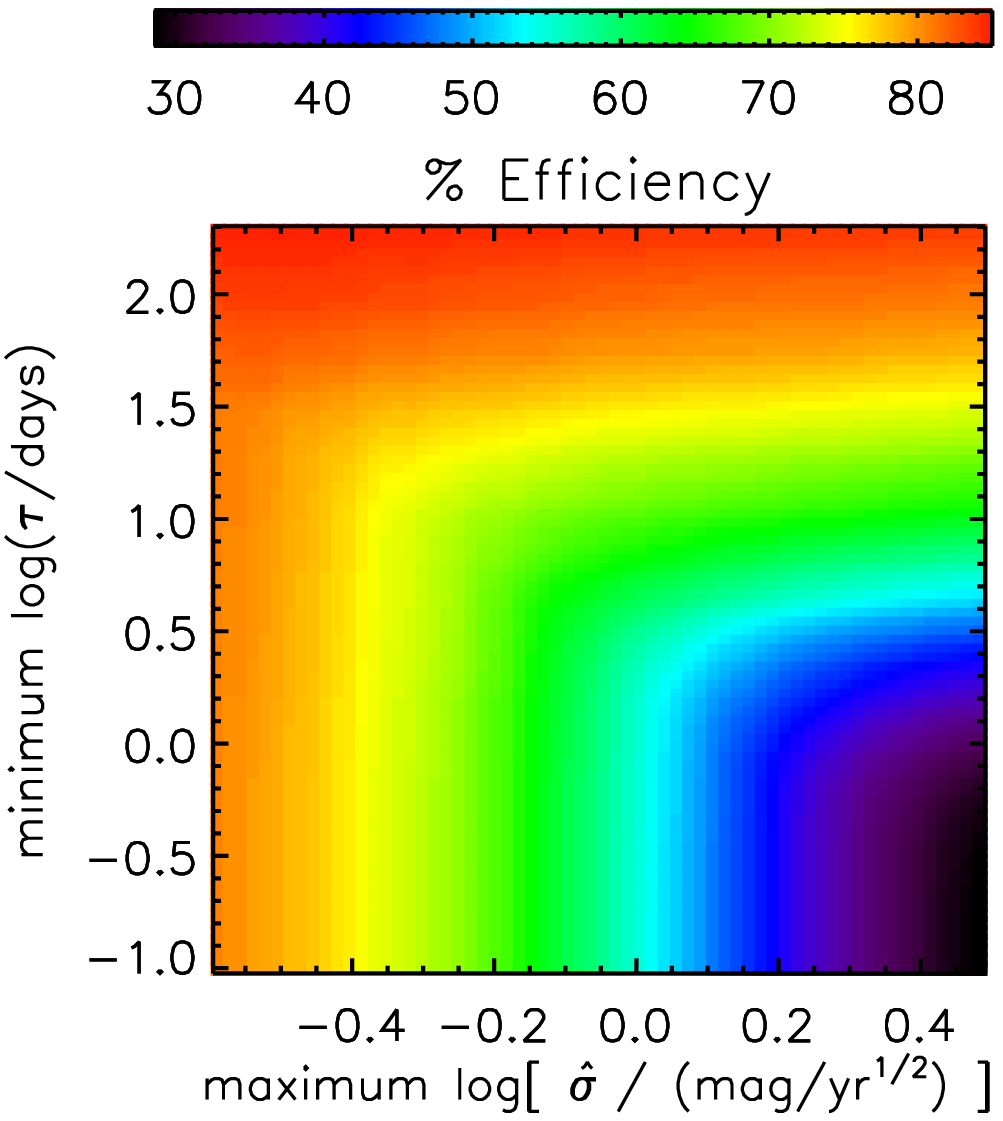}
\includegraphics[width=2in]{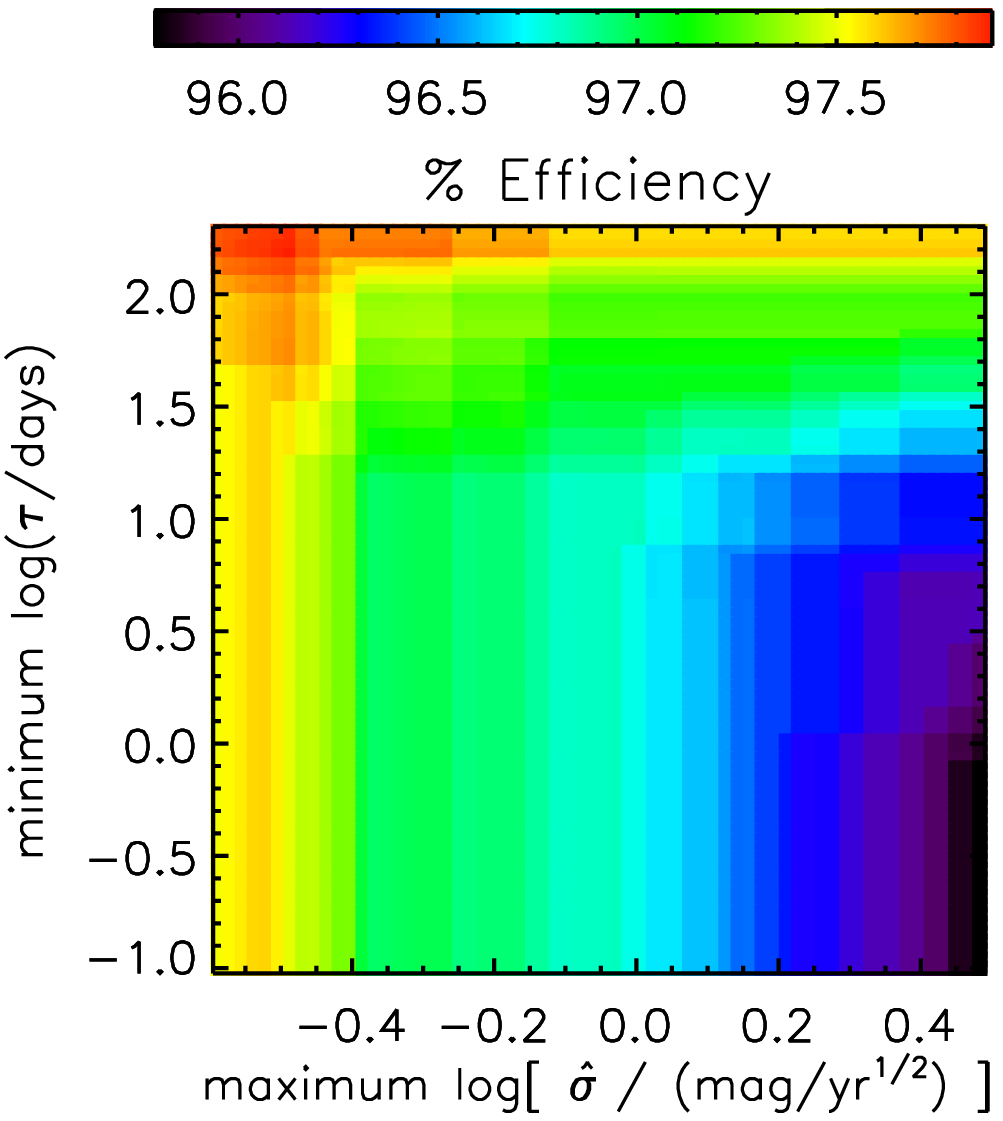}}\centerline{
\includegraphics[width=2in]{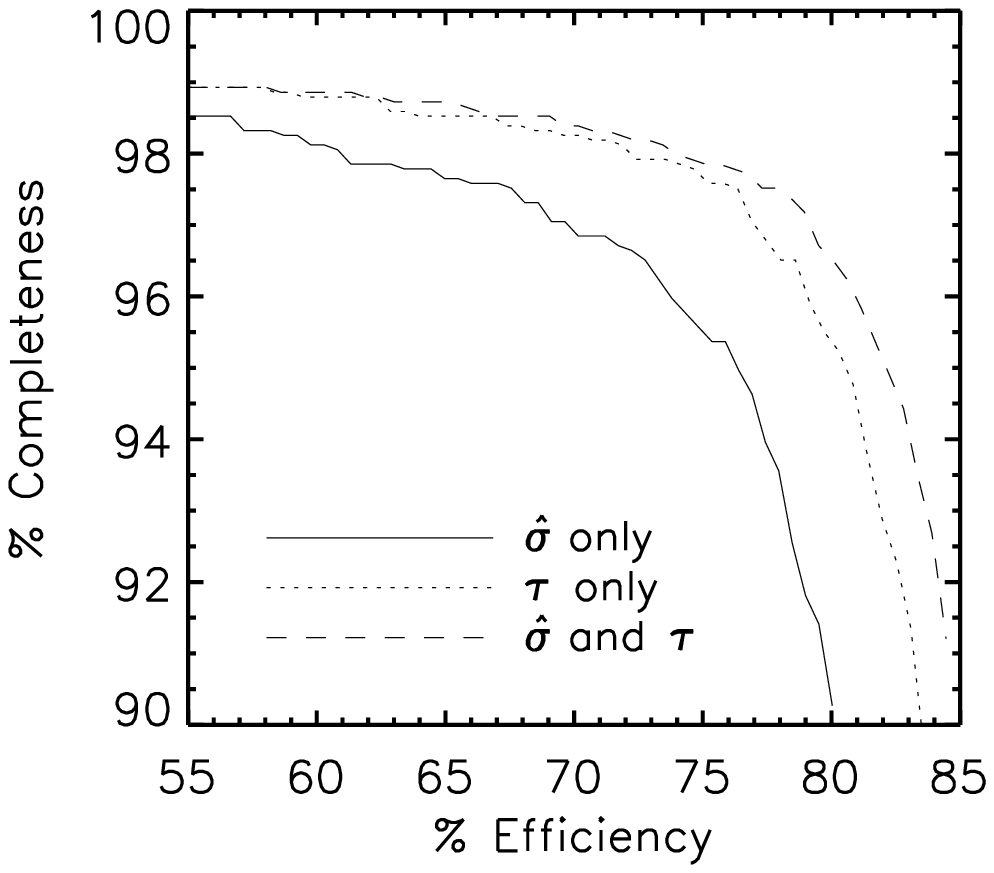}
\includegraphics[width=2in]{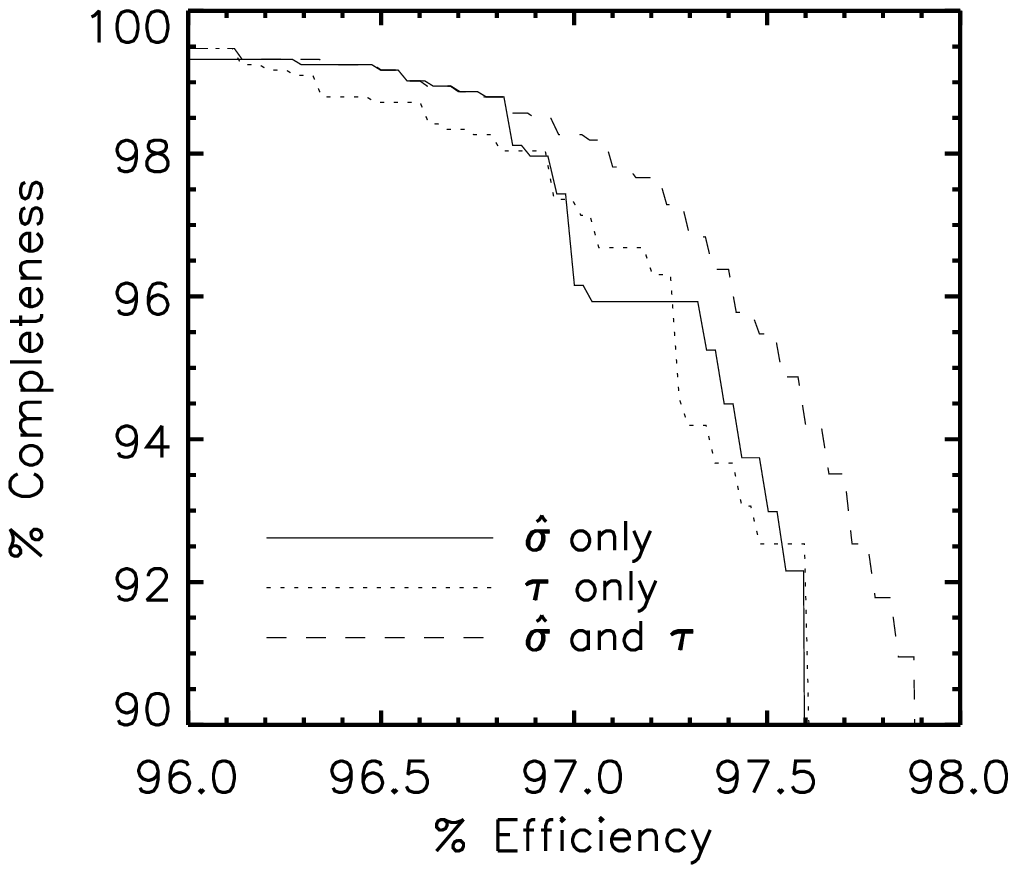}}
\caption{\footnotesize  \emph{Left panels:} 
  Improvements in variability-based selection of 
  quasars due to added time scale information. 
  The top two panels show the the completeness ($C$) and
  efficiency ($E$) for quasar classification as a function of the
  minimum $\tau$ and maximum $\hat{\sigma}$ allowed for the variable 
  sample with $i<19$ and $-35\,^{\circ}<~$RA$~<~50\,^{\circ}$.  
  The solid line in the bottom panel shows 
  the maximum $C$ as a function of $E$ 
  achieved when using $\hat{\sigma}$ information 
  alone (i.e., allowing all time scales $\tau$). 
  The dashed line shows how $E$ improves 
  when including a cut in $\tau$ to select quasars in addition to a
  cut in $\hat{\sigma}$. The dotted line shows results when 
  using the $\tau$ information alone. 
  \emph{Right panels:} As in left panels but for UVX objects only. 
} 
\label{fig:optimaltausig}
\end{figure*}

The driving amplitude $\hat{\sigma}$ also shows a large contrast
between the quasar color regions (II and IV) and the stellar regions
(I and III) in Figure~\ref{fig:colormap}. This suggests that including
information on the driving amplitude may improve the quasar selection
based on variability. Indeed, Schm10 found high-efficiency quasar
samples by using information closely corresponding to $\hat{\sigma}$. 
In Figure~\ref{fig:optimaltausig}, we show how the selection can be
improved by including information on both $\tau$ and $\hat{\sigma}$.  
In the top two panels, $E$ and $C$ are mapped out as a function of the 
minimum $\tau$ and the maximum $\hat{\sigma}$ allowed for quasar
classification.  For every value of $\hat{\sigma}$, there is 
a lower limit on $\tau$ that boosts efficiency with a minor drop in 
$C$. This essentially follows one of the quasar selection
criteria for $\tau$ and $\hat{\sigma}$ in Koz{\l}10. A similar
improvement is seen when instead using lower limits in both $\tau$ and
$\rm{SF}_{\infty}$, or a combination of $\tau$, $\rm{SF}_{\infty}$,
and $\hat{\sigma}$ information (the differences are all $\lesssim 1$\%). 
In the bottom panels of Figure~\ref{fig:optimaltausig}, it can be seen 
more quantitatively how the inclusion of $\tau$ improves the
selection of quasars as compared to using $\hat{\sigma}$ alone.  
The panels show the maximum $C$ achieved for
an $E$ exceeding that plotted on the x-axis, for three cases. In
the first case (solid), only the $\hat{\sigma}$ parameter is used to
separate quasars and stars, i.e., all values of $\tau$ are allowed, as
in Schm10. The second case (dashed) shows the effect of including
information on $\tau$. The third case (dotted) shows results for using
$\tau$ information alone.  When including all SDSS colors (bottom-left
panel), the $\tau$ information alone leads to significantly higher $E$
than a selection based on $\hat{\sigma}$ alone. However, for UVX
objects (bottom-right panel), a combination of $\tau$ and
$\hat{\sigma}$  information is required to significantly boost $E$
from a selection based on $\hat{\sigma}$ alone.

\begin{figure*}[h!]
\epsscale{.7}
\plotone{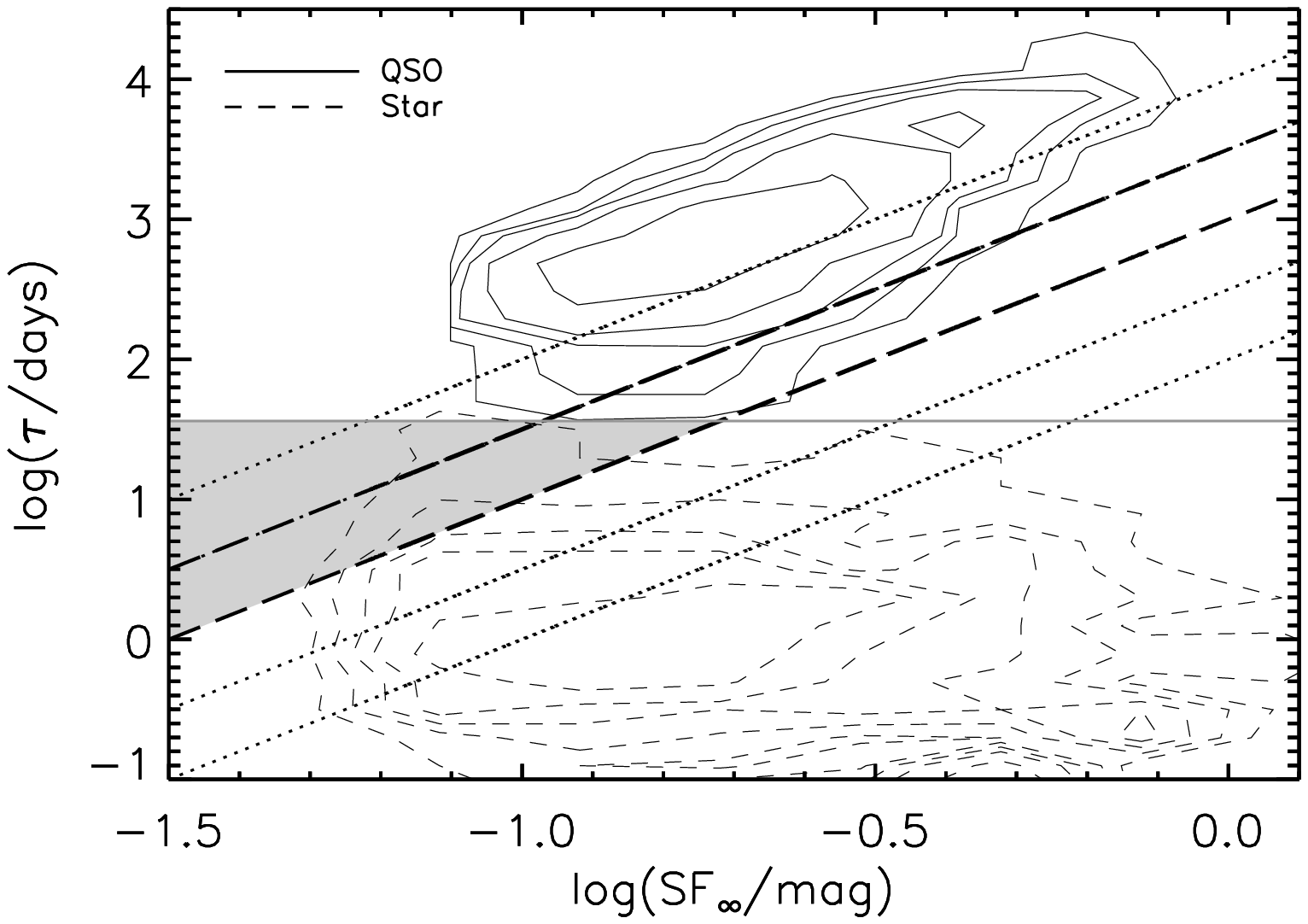}
\plotone{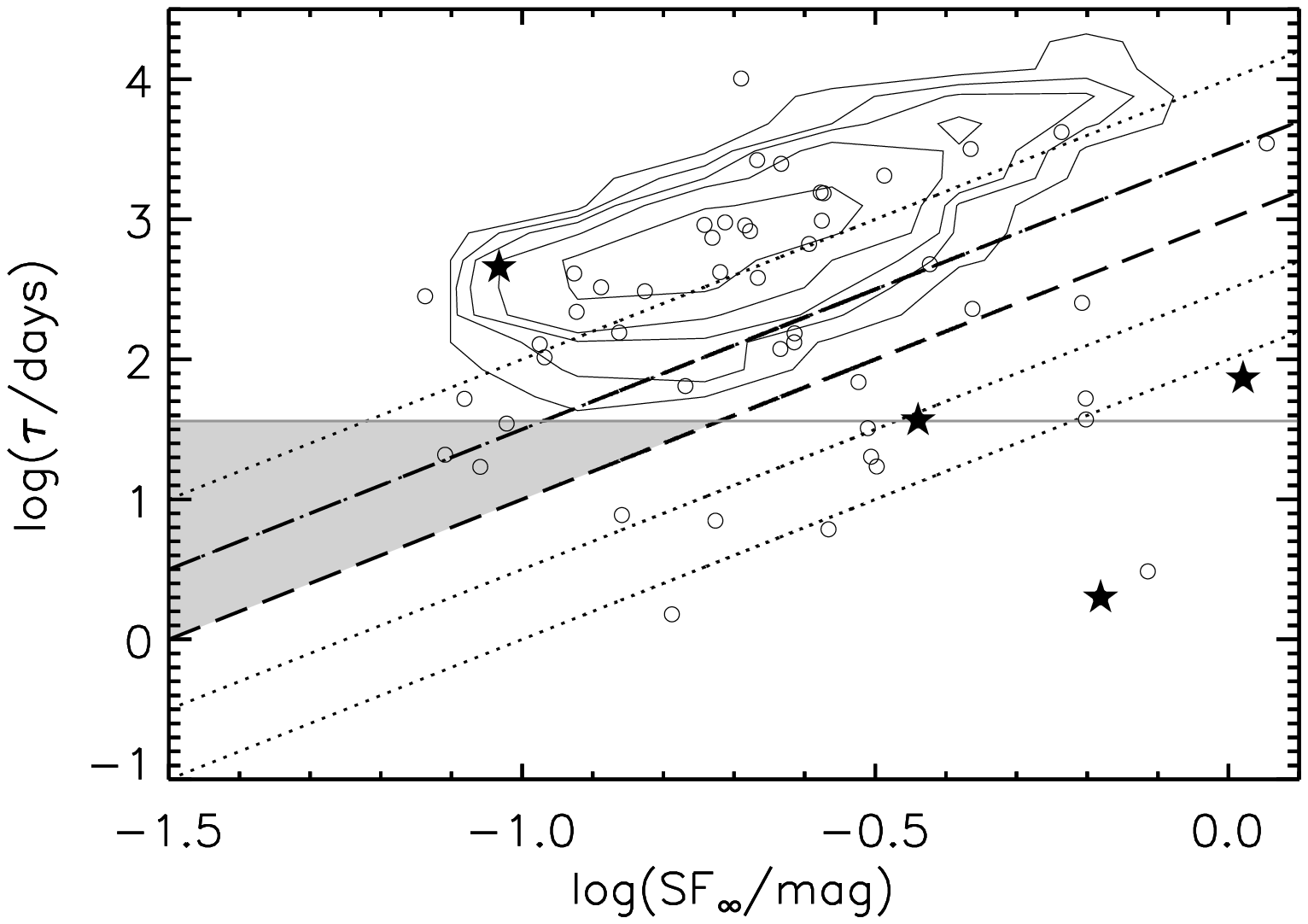}
\caption{\footnotesize 
  \emph{Top:} Characteristic time scale $\tau$ versus asymptotic rms 
  variability $\rm{SF}_\infty$ for  confirmed 
  quasars (solid) and stars (dashed) in our
  S82 data set restricted to $i<19$. The
  contours show regions containing 40, 60, 75, 85, and 90\% of data
  points in each subsample. 
  The dotted lines represent lines of constant
  $\hat{\sigma}=\rm{SF}_{\infty}/\sqrt{\tau}$. 
  The two thick dashed lines correspond to $\hat{\sigma}=10^{-0.22}$ and
  $10^{-0.47}$ mag yr$^{-1/2}$. The gray region represents the
  contaminating (stellar) region when selecting sources with
  $\hat{\sigma}<10^{-0.22}$ mag yr$^{-1/2}$.  When imposing a lower
  limit at $\tau = 10^{1.56}$~days (gray horizontal line),
  these contaminants are excluded from the sample, leading to a higher
  efficiency of quasar selection.  \emph{Bottom:} As in top panel but
  for UVX objects only. Here, the confirmed stars
  are shown with filled stars, and UVX objects without SDSS spectra
  are shown with open circles.
}
\label{fig:sftausig}
\end{figure*}

Using both $\tau$ and $\hat{\sigma}$ as criteria for quasar selection
in the absence of color information, one can achieve an efficiency of
$E=82$\% with a completeness of $C=96$\% (seventh row of
Table~\ref{tab:cuts}), or alternatively, $E=75$\% for $C=98$\%, and 
$E=85$\% for $C=90$\%.  Here, objects that drop out 
of the selection after omitting the most outlying data point in the light 
curve are excluded.  When restricting the selection to color regions
II and IV in Figure~\ref{fig:colormap}, one can achieve $E=96$\% with
$C=93$\%.  At the same level of completeness, $E=84$\% without color 
cuts, and $E=78$\% when using $\hat{\sigma}$ alone. 
For UVX objects, $E=97$\% and $C=98$\% when selecting 
$\tau \geq 10^{1.56}$~days and $\hat{\sigma} \leq 10^{-0.18}$ mag
yr$^{-1/2}$ (ninth row of Table~\ref{tab:cuts}). For illustration, 
Figure~\ref{fig:sftausig} shows $\tau$ versus $\rm{SF}_\infty$ for
the variable S82 sample, with lines of constant $\hat{\sigma}$
overlaid. By choosing $\hat{\sigma}<10^{-0.2}$ mag yr$^{-1/2}$, the
entire quasar locus is selected as well as the gray shaded region
containing stars. However, when imposing a lower limit at 
$\tau = 10^{1.56}$~days, these contaminants are excluded from the sample,
leading to a higher efficiency of quasar selection.

Since the selection of quasars based on variability is a useful tool
in regions of high stellar density (Koz{\l}10), we repeat the analysis in
Table~\ref{tab:cuts} for the region
$-35\,^{\circ}<$~RA~$<-25\,^{\circ}$, where the density of Galactic
stars rises. Since the rate of contamination is higher in this region, 
it is useful to reject outliers in light curves in order to boost
efficiency as in the fourth row of Table~\ref{tab:cuts}. These results
can be found in Table~\ref{tab:cuts2}. Here, $E$ for the 
$\tau \geq 100$~days selection is 79\%, after omitting outliers in light curves. 
Restricting this selection to $\Delta L_{\rm{noise}}>10$ boosts $E$ to 82\%
without a drop in completeness.  
Therefore, a strict cut in $\Delta L_{\rm{noise}}$ is especially useful in
regions of high stellar density.

\begin{deluxetable}{c c c c c c c  c c c c c}
\tabletypesize{\footnotesize}
\tablewidth{0pt}
\tablecaption{Selection Criteria for $-35\,^{\circ}<$~RA~$<50\,^{\circ}$ and $i<19$\label{tab:cuts}}
\tablehead{
& Selection                              & $N$ & $C$ & $\delta C$ & $E$ & $\delta E$ & Spec.\ & QSO & Star & Gal.\ & Unk.\ 
}
\startdata
1.\ & All Sources                                                          & 10024 & 100 &  4 & 15 &0.4 & 33  &  45 & 53& 2.5$\phantom{0}$  & 0.15 \\
\hline				       	                              			                        	     
2.\ & $\tau$ \textgreater 1 d                                              & 5385  & 99.7&  4 & 28 &0.8 & 43  &  64 & 33& 2.0$\phantom{0}$  & 0.22 \\
3.\ & $\tau$ $\geq$ 100 d                                                  & 1734  & 94  &  3 & 81 &  3 & 84  &  96 & 3.5$\phantom{0}$& 0.55 & 0.14 \\
4.\ & $\tau$ $\geq$ 100 d\tablenotemark{*}                                 & 1594  & 93  &  3 & 87 &  3 & 89  &  98 & 1.76 & 0.49 & 0.14 \\
5.\ & $\tau$ $\geq 10^{1.5}$~d, $\Delta L_{\rm{noise}}>10$\tablenotemark{*}&  1647 &  96 &  4 & 87 &  3 & 89  &  97 &  2.0$\phantom{0}$&  0.61 & 0.20\\
6.\ & $\tau$ $\geq$ 100 d, qso-like colors                                 & 1420  & 91  &  3 & 96 &  4 & 97  &  99 & 0.15 & 0.44 & 0.15 \\
7.\ & $\tau$ $\geq 10^{1.56}$~d, $\hat{\sigma} \leq 10^{-0.18}$ mag			                        	     
yr$^{-1/2}$\tablenotemark{*}                                               &  1751 &  96 &  4 & 82 &  3 & 86  &  96  & 3.3$\phantom{0}$ & 0.87 & 0.07\\
\hline				           	  					                        	     
8.\ & {\bf UVX:}~~~~~~ $\tau$ $\geq$ 100 d~~~~~~~~~~~~~~~~                 & 1297  & 95  &  3 & 97 &  4 & 98  &  99 & 0.16 & 0.32 & 0.16 \\
9.\ & $\tau$ $\geq 10^{1.56}$~d, $\hat{\sigma} \leq 10^{-0.18}$ mag			                        	     
yr$^{-1/2}$                                                                & 1332  &  98 &  3 & 97 &  4 & 98  &  99 &  0.15&  0.31&  0.08 \\
\hline				           	  					                        	     
10.\ & {\bf nUVX:}~~~~~ $\tau$ $\geq$ 100 d~~~~~~~~~~~~~~~~	           & 52    & 100 &0.4 & 69 & 15 & 69  & 100 & 0.00 & 0.00 & 0.00 \\
\hline				           	  					                        	     
11.\ &  B\&B                                                               & 1702  & 96  &  4 & 84 &  3 & 87  &  97 &  2.7$\phantom{0}$&  0.40&  0.13\\
\tableline
\enddata
\tablenotetext{*}{The most outlying data point 
  in each light curve is omitted.}
\tablecomments{ The first column lists various selection
  criteria (``qso-like colors'' signifies having
  $u-g$ and $g-r$ colors in regions II or IV of
  Figure~\ref{fig:colormap}).  $N$ is the total number of objects that satisfy each criterion (the
  sample is defined by $i<19$ and covers 213~deg$^2$). $C$ is 
  the quasar completeness, or the percentage
  of all confirmed quasars in the sample of 10,024 
  that are selected.  $E$ is the efficiency, or the percentage
  of the selected objects that are confirmed
  quasars. $\delta C$ and $\delta E$ are the Poisson errors on $C$ and $E$. 
  ``Spec.'' indicates the percentage that have an
  SDSS spectrum, and the last four columns show the percentage breakdown 
  of the subset with spectra into four spectroscopic classes: QSO
  (SpecClass is 3 or 4), Star (SpecClass is 1 or 6), Galaxy (SpecClass
  is 2), and Unknown (SpecClass is 0).  In the eighth and ninth rows (tenth row), 
  we list results for the (n)UVX color box described in Schm10,
  where $C$ is computed as the percentage of all
  confirmed quasars contained in the (n)UVX color box
  that satisfy the variability criteria listed in the first column. 
  The eleventh row lists the results for the selection criteria in B\&B
  (see Section~\ref{sec:compareBB}).
}
\end{deluxetable}

\begin{deluxetable}{c c c c c c c  c c c c c}
\tabletypesize{\footnotesize}
\tablewidth{0pt}
\tablecaption{Selection Criteria for $-35\,^{\circ}<$~RA~$<-25\,^{\circ}$ and $i<19$ \label{tab:cuts2}}
\tablehead{
& Selection                                  & $N$ & $C$ & $\delta C$ & $E$ & $\delta E$ & Spec.\ & QSO & Star & Gal.\ & Unk.\ 
}
\startdata
1.\ & All Sources                                                     &2298 &  100& 12& 6.5 &0.5 &20 & 32 & 65 & 2.17 & 0.22 \\ 
\hline										    	      		     
2.\ & $\tau$ \textgreater 1 d                                         & 991 &  99 & 12& 15  & 1  &30 & 50 & 47 & 2.72 & 0.34 \\
3.\ & $\tau$ $\geq$ 100 d\tablenotemark{*}                            & 167 &  89 & 11& 79  & 9  &83 & 95 & 2.2$\phantom{0}$ & 2.16 & 0.72\\ 
4.\ & $\tau$ $\geq$ 100 d, $\Delta L_{\rm{noise}}>10$\tablenotemark{*}& 160 &  89 & 11& 82  & 10 &85 & 97 & 0.74 & 1.47 & 0.74\\
5.\ & $\tau$ $\geq$ 100 d, qso-like colors\tablenotemark{*}           & 137 &  86 & 10& 93  & 11 &96 & 98 & 0.76 & 1.53 & 0.76 \\
6.\ & $\tau$ $\geq 10^{1.56}$~d, $\hat{\sigma} \leq 10^{-0.18}$ mag		    	     		     
yr$^{-1/2}$,                                                          & 179 &  93 & 11& 78  & 9  & 80 & 97 &  1.4$\phantom{0}$&  1.4$\phantom{0}$&  0.70\\
    & ~~~~~$\Delta L_{\rm{noise}}>10$\tablenotemark{*}                &     &     &   &     &    &    &    &     &       & \\
\hline										    	     		     
7.\ & {\bf UVX:}~~~~~~ $\tau$ $\geq$ 100 d~~~~~~~~~~~~~~~~            & 127 &  89 & 10& 94  & 12 & 97 & 98 & 0.00 & 1.63 & 0.81 \\
8.\ & $\tau$ $\geq 10^{1.56}$~d, $\hat{\sigma} \leq 10^{-0.18}$ mag		    	     		     
yr$^{-1/2}$                                                           & 134 &  95 & 10& 96  & 12 & 97 & 98 &  0.00 & 1.54&  0.00\\
\hline				           	  				    	     		     
9.\ & {\bf nUVX:}~~~~~ $\tau$ $\geq$ 100 d~~~~~~~~~~~~~~~~            & 10  & 100 & 2 & 50  & 27 & 50 & 100 & 0.00 & 0.00 & 0.00\\
\tableline
\enddata
\tablenotetext{*}{The most outlying data point 
  in each light curve has been omitted.}
\tablecomments{ As in Table~\ref{tab:cuts} but for regions of increased stellar density.}
\end{deluxetable}

\clearpage
\begin{figure*}[t!]
\epsscale{.5}
\plotone{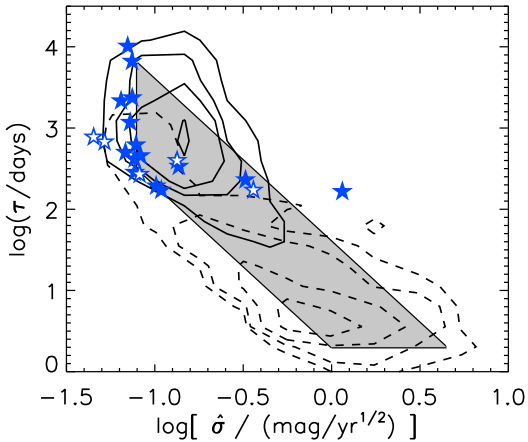} 
\caption{\footnotesize  Variability time scale $\tau$ as a function of short-term
 driving amplitude $\hat{\sigma}$. The contours show the regions
 containing 25, 50, 75, and 90\% of data points for quasars
 (solid, $N = 1170$) and all other objects (dashed, 
 $N = 1783$). The shaded region shows the quasar selection cut from
 Koz{\l}10. The star symbols represent the confirmed
 stars passing our $\tau > 100$~days and $\Delta L_{\rm{noise}}>2$
 selection criteria (except those flagged as having outliers in their
 light curves -- see Table~\ref{tab:targets} caption). The star symbols
 filled with white indicate late-type stars (SDSS SpecClass$=6$). 
} 
\label{fig:sigtau}
\end{figure*}

\subsubsection{Comparison to Koz{\l}owski et al.~(2010)}
\label{sec:compareKozl}
Koz{\l}10 proposed a quasar variability selection regions based on the light 
curves of OGLE-II and OGLE-III sources in the Magellanic Clouds with the 
mid-IR colors of AGN from Koz{\l}owski \& Kochanek (2009) also with the
requirement that $\Delta L_{\rm{noise}}>2$.  Figure~\ref{fig:sigtau} shows the distribution 
of spectroscopically identified quasars and other sources in this selection
space as well as the confirmed stars passing the $\tau > 100$~days and 
$\Delta L_{\rm{noise}}>2$ selection cuts.  The completeness 
of this selection region for the SDSS quasar sample is relatively
low, $C=55\%$, as is the efficiency, $E=46\%$ (62\% when excluding
outliers in light curves), which demands some
explanation.  The lower completeness is likely a consequence of the
differences in the cadence and duration of the OGLE light curves as 
compared to the SDSS light curves.  The OGLE light curves are densely
sampled over the full durations of the surveys, and the typical OGLE-III
light curve has 370 epochs over 7 years.  As a result, fewer light
curves have run-away estimates of $\tau$, making it relatively safe
to impose an upper limit on $\tau$ in order to reduce contamination by
long period variables (LPVs).  In the very dense stellar environment
of the Magellanic Clouds, the contamination by LPVs is significant.
For the SDSS regions, if we simply eliminate the upper boundary of the 
Koz{\l}10 selection region, the completeness rises to $C=91\%$ with 
a corresponding rise in the efficiency to $E=52\%$ (69\%, excluding outliers) because
many more quasars are added than contaminating stars.  Figure~\ref{fig:sigtau}
does show significant contamination coming from sources with
short time scales, and this was not observed in the Magellanic
Cloud sample after removing the periodic variables (predominantly
RR~Lyrae and Cepheids).  Part of the problem may again be the
differences in the light curve cadences, where the better cadenced
OGLE light curves can better distinguish $\tau \equiv 0$ from a small 
$\tau$, leading to less ``leakage'' of false sources past the
$\Delta L_{\rm{noise}}>2$ boundary.  There are probably also some
basic differences in the sources of contamination.  Koz{\l}10 were
designing selection criterion for a small area ($1.3$~deg$^2$) containing
a large number of luminous stars whose natural time scales are
relatively long.  Here, we are examining a much wider area
(213~deg$^2$) where the variable stars are generally not as luminous
and/or are white dwarfs, leading to relatively shorter natural time scales.
Thus, the generous boundary for short $\tau$ in Koz{\l}10 may not
be appropriate to a wide area survey.  If we raise the $\tau >2$~day
lower limit in Koz{\l}10 to $\tau > 10^{1.5}$~days, then for the SDSS region the
completeness changes little ($C=90\%$), but the efficiency rises
dramatically to $E=78\%$ (86\%, excluding outliers).  Finally, if we consider the variability
properties of the variable spectroscopically identified stars
in this diagram, they tend to lie in the region of low amplitudes
$\hat{\sigma}$ and long time scales. In the Magellanic Cloud
regions, where Koz{\l}10 could take advantage of color magnitude diagrams 
to better characterize the variable stars, these tend to be various
classes of evolved, red stars -- long period variables (LPV),
long secondary period (LSP) variables and OGLE Small Amplitude Variable 
Red Giants (OSARGS).

\subsubsection{Comparison to Butler \& Bloom (2010)}
\label{sec:compareBB}
Butler \& Bloom (2010, hereafter B\&B) present a different approach to
using the DRW model to select quasars based on variability. They model
the ensemble SF for quasars (see, e.g.,
Vanden Berk et al.~2004) as a function of apparent magnitude with the
DRW model, and use this to predict $\tau$ and $\hat{\sigma}$ for
individual quasar candidates. 
The process model $\chi^2$ is then computed for each light
curve using these {\it fixed} DRW parameters to obtain $\chi^2_{\rm{QSO}}$. 
Essentially, they fit the DRW model with a (infinitely) strong prior
on the values of $\tau$ and $\hat{\sigma}$ derived from the apparent
magnitude. They also compute $\chi^2_{\rm{FALSE}}$, the goodness of fit
expected if the same DRW model was fit to a light curve that was really
white noise with the overall variance of the observed light curve. 
The main difference with respect to our method is that we obtain
individual estimates of $\tau$ and $\hat{\sigma}$ for every light
curve independent of the mean magnitude. This leads to a much
wider distribution of $\tau$ and $\hat{\sigma}$ for our method with
respect to B\&B, allowing us to use $\tau$ as an efficient classifier.
The selection technique used in B\&B is very efficient, however,
because only a single model rather than a parameter fit is required. 

Here we compare the resulting $C$ and $E$ from each method using
identical samples of light curves. 
In Table~\ref{tab:cuts} we compare the two methods for the light curve
sample with $-35\,^{\circ}<$~RA~$<50\,^{\circ}$ and $i<19$, where the
B\&B method selects an object as a quasar candidate if
\begin{itemize}
\item $\chi^2_{\rm{FALSE}}/\nu > \chi^2_{\rm{QSO}}/\nu$
\item $\chi^2_{\rm{FALSE}}/\nu > 3.1$ if RA~$>-20$~deg (5$\sigma$ significance)
\item $\chi^2_{\rm{FALSE}}/\nu > 4.4$ if RA~$<-20$~deg (7$\sigma$ significance),
\end{itemize}
where $\nu$ is the number of degrees of freedom.  Comparing the fourth
and eleventh rows, one can see that $C$ is higher by a few percent for
the B\&B method whereas $E$ is slightly lower. When instead choosing 
$\tau>10^{1.5}$~days and $\Delta L_{\rm{noise}}>10$, a similar completeness
is achieved (96\%) with $E=87$\%.  Therefore, with some fine-tuning of
parameters, the two methods can deliver samples with comparable $C$
and $E$, which is not too surprising since they rely on the same
mathematical model to describe the light curve behavior.  
The B\&B method is faster if the only goal is to select quasars, but
fitting for $\tau$ and $\hat{\sigma}$ then allows for further
classification of the type of variability, as illustrated by the
locations of different classes of variable stars in the parameter
space (see Figure~9 in Koz{\l}10). 

In Table~\ref{tab:cross}, we present the cross--correlation 
between our method and that in B\&B in terms of how many $i<19$
sources are selected using either technique. For this table, we
define our selection criteria as $\tau \geq 10^{1.5}$~days and
$\Delta L_{\rm{noise}}>10$ (after omitting light curve outliers). We define
the B\&B selection criteria as listed above. We then show the
characteristics of the samples defined by passing both criteria, at
least one of the criteria, failing both, and passing only one or the other.
It is beneficial to use both methods to select
quasars. One can achieve $E=90$\% in the absence of color selection
when requiring both criteria, but $C=97$\% can be maintained when
using either criteria for an $E$ of 82\%. In essence, some quasars
have true $\tau$ and $\hat{\sigma}$ values too far from the prior
defined by the apparent magnitude and so have reduced B\&B
likelihoods, while for some other objects, the addition of the strong B\&B prior
leads to better rejection of false positives and negatives.

\begin{deluxetable}{c c c c c c c  c c c c c}
\tabletypesize{\footnotesize}
\tablewidth{0pt}
\tablecaption{Cross--Correlation with B\&B for $-35\,^{\circ}<$~RA~$<50\,^{\circ}$ and $i<19$ \label{tab:cross}}
\tablehead{
& Selected By  & $N$ & $C$  & $\delta C$ & $E$ & $\delta E$ & Spec.\ & QSO & Star & Gal.\ & Unk.\ 
}
\startdata
1.\ & All Sources      & 10024 & 100 & 4   & 15  & 0.4 & 33 &  45 & 53              & 2.5$\phantom{0}$ & 0.15 \\
2.\ & Both Methods     & 1564  & 95  & 3   & 90  & 3   & 92 &  98 & 1.3$\phantom{0}$& 0.35             & 0.14 \\
3.\ & Either Method    & 1763  & 97  & 4   & 82  & 3   & 85 &  96 & 3.1$\phantom{0}$& 0.60             & 0.13 \\
4.\ & Neither Method   & 8237  & 2.6 & 0.4 & 0.5 & 0.1 & 22 & 2.1 & 94              & 4.0$\phantom{0}$ & 0.11 \\
5.\ & Only our Method  &  83   & 0.9 & 0.3 & 17  & 5   & 37 &  45 & 39              & 13               & 3.2$\phantom{0}$ \\
6.\ & Only B\&B        & 138   & 1.5 & 0.3 & 16  & 4   & 33 &  49 & 49              & 2.2$\phantom{0}$ & 2.2$\phantom{0}$ \\
\tableline
\enddata
\tablecomments{Column definitions follow those in Table~\ref{tab:cuts}. The
  second row lists results for a sample satisfying both the B\&B
  criteria as well as our criteria ($\tau \geq 10^{1.5}$~days and
  $\Delta L_{\rm{noise}}>10$, after omitting light curve outliers). The
  third row lists results for a sample satisfying either the B\&B
  criteria or ours. The selection in the 
  fourth row is rejected by both methods, while the selections in the
  fifth and sixth rows only satisfy one of the methods.}
\end{deluxetable}

\section{Effect of Cadence and Photometric Accuracy on Best-fit
  Variability Parameters: Prospects for LSST and Pan-STARRS}
\label{sec:lsst}

One clear lesson from our comparison of our proposed SDSS selection
criteria (applied to sparsely sampled, effectively shorter duration light curves in 
a wide area, extragalactic field) and the Koz{\l}10 selection criteria
(applied to densely sampled, longer duration light curves in a small area with
high stellar densities) is that the appropriate selection criteria
will depend on both the properties of the light curves (duration,
cadencing, and uncertainties) and the nature of the contamination.  
Here we consider the effects that the survey length, time sampling, and
photometric errors have on the best-fit DRW parameters. These
are important issues to consider for upcoming surveys such as
Pan-STARRS (Kaiser et al.~2002), which is already taking data, and
LSST (Ivezi\'{c} et al.~2008), which will obtain accurate, 
well-sampled light curves for millions of quasars. To probe the time
scales as short as $0.1\tau$, and assuming a rest-frame time scale of
$\tau=10$~days and a characteristic redshift of 2, the light curves
should be sampled every 3 days in the observer's frame, which is in
good agreement with the baseline cadence of LSST.  
The LSST photometric errors in the $r$ band will be $<0.02$
mag for $r<22$, and there are roughly 2-3 million AGN with $r<22$ in
the 20,000 sq.\ deg.\ covered by the main LSST survey (see Table
10.2 in the LSST Science Book; LSST Science Collaborations and LSST Project 2009). 
Each of these objects will be observed about 1000 times (summed over
all bands), yielding a database of over 2 billion photometric
measurements. The characteristic time scale $\tau$ will be 
well-constrained if the survey length is sufficiently longer than
$\tau$. It is therefore necessary to evaluate the accuracy of fitted
variability parameters as a function of light curve length. 

\subsection{The Impact of Light Curve Length}
As a first test, we generate 10,000 DRW light curves
using an input of $\tau=575$~days and $\rm{SF}_{\infty}=0.2$ mag, which are the
median observed-frame values for the S82 quasar sample. The light
curves are sampled once every 10 days ($\Delta t=10$~days) and span a length of 40 years. 
Photometric errors are drawn from a Gaussian distribution with a
standard deviation of $\sigma=0.01$ mag. To test the impact of light
curve length, we truncate the light curves to 10 years, 3 years, and 1
year and compare the resulting variability parameters. For data
lengths shorter than $\tau$, only the driving amplitude of short-term
fluctuations, $\hat{\sigma}=\rm{SF}_{\infty}/\sqrt{\tau}$, is well 
constrained [recall that $SF(\Delta t << \tau) =
\hat{\sigma}\sqrt{|\Delta t|}$]. That is, on a grid of $\log(\tau)$ versus
$\log(\rm{SF}_{\infty})$, the best-fit values will be scattered due to
fitting errors along lines of constant $\hat{\sigma}$, but much less
perpendicular to it (i.e., along lines of constant
$K=\tau\sqrt{\rm{SF}_{\infty}}$). Therefore, as the light curve length
decreases, the mean best-fit $\hat{\sigma}$ will not vary
significantly, while the best-fit parameters $\tau$ and $\rm{SF}_{\infty}$,
and $K$ will become biased (see Koz{\l}10; MacLeod et al.~2010). Note that
only two of these four parameters are independent. 

\begin{figure}[h!]
\epsscale{.61}
\plotone{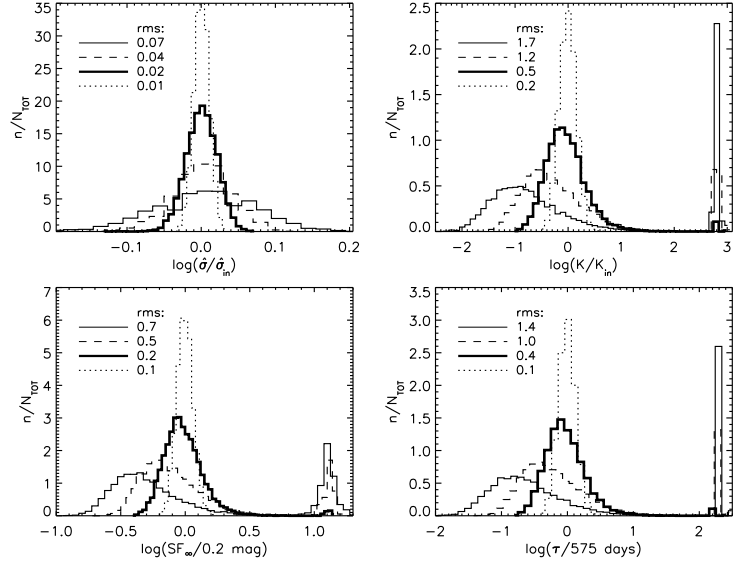}
\caption{\footnotesize Distributions of  
  $\hat{\sigma}=\rm{SF}_{\infty}/\sqrt{\tau}$, $K=\tau\sqrt{\rm{SF}_{\infty}}$,
  $\tau$, and $\rm{SF}_{\infty}$ normalized by the input values for
  10,000 simulated light curves with length 40 years (dotted), 10 years
  (thick), 3 years (dashed), and 1 year (thin solid), all using 
  $\Delta t=10$~days and photometric accuracy of 0.01 mag. 
  $n$ indicates the number of points in a bin divided by the bin width, and 
  $N_{TOT}$ is the total number of points used for each histogram. 
  The rms of each distribution is listed in the legend.}
\label{fig:length}
\end{figure}

Figure~\ref{fig:length} shows the distribution of best-fit variability
parameters $\hat{\sigma}$, $K$, $\tau$, and $\rm{SF}_{\infty}$ normalized
by the input values for simulated light curves with length 40 years,
as well as when truncated to 10 years, 3 years, and 1 year. As the
length increases, the widths of the output parameter distributions
decrease, roughly as the square root of the light curve length. While
the mean output value of $\hat{\sigma}$ is unaffected by the light
curve length, the mean values of $K$, $\tau$, and $\rm{SF}_{\infty}$ 
decrease as the length shortens, and the number of ``run-away'' time
scales ($\tau \geq 10^5$~days) increases due to the fact that $\tau$
is now longer than the light curve length and cannot be
constrained. We find that for typical variable quasars, the 
best-fit distributions of $\tau$ and $\rm{SF}_{\infty}$ are strongly biased 
for surveys significantly shorter than 10 years (by about 70\% and 40\%, 
respectively, for a length of 3 years). Nevertheless, the
$\hat{\sigma}$ parameter is unbiased and already well-measured
($\rm{rms}=20$\%) after 1 year. 

\subsection{The Impact of Realistic Cadence}
\begin{figure}[h!]
\epsscale{.7}
\plotone{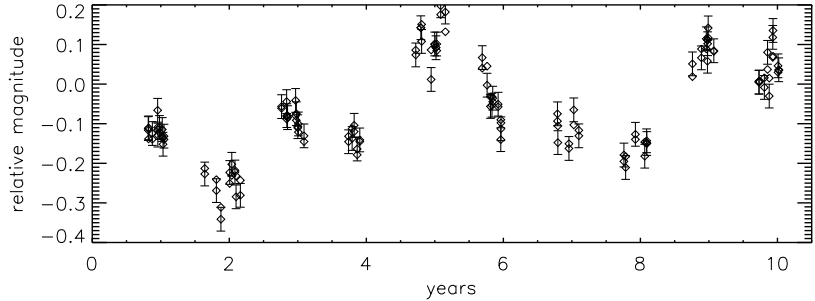}
\caption{\footnotesize A model light curve generated for  $\tau=575$~days, 
 $\rm{SF}_{\infty}=0.2$ mag, the $r$-band LSST cadence, and a photometric
 accuracy of 0.03 mag. 
 }
\label{fig:LCegLSST}
\end{figure}

\begin{figure}[h!]
\epsscale{.61}
\plotone{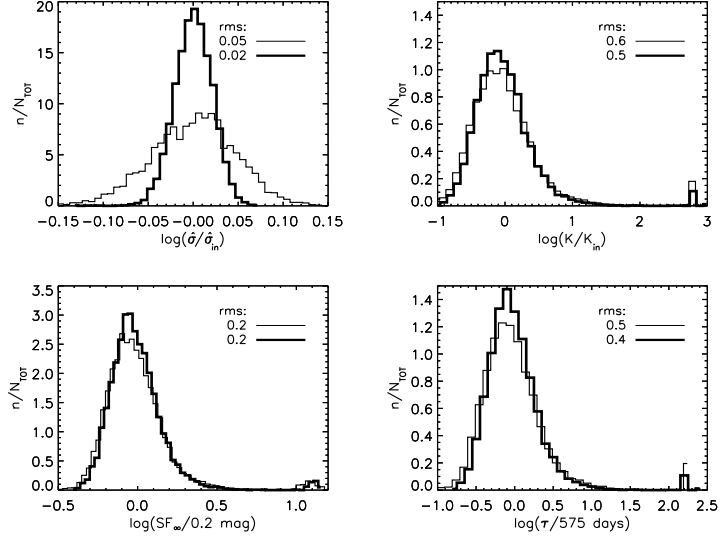}
\caption{\footnotesize Distributions of  
  $\hat{\sigma}=\rm{SF}_{\infty}/\sqrt{\tau}$, $K=\tau\sqrt{\rm{SF}_{\infty}}$,
  $\tau$, and $\rm{SF}_{\infty}$ normalized by the input values for 10,000 
  simulated light curves with a photometric accuracy of 0.01 mag and
  the $r$-band LSST cadence with $\sim$200 data points (thin solid). 
  The thick line shows the distribution from
  Figure~\ref{fig:length} with a cadence of $\Delta t=10$~days over 10
  years (365 data points). The rms of each distribution is listed in
  the legend.  
}
\label{fig:cadence}
\end{figure}

As a second test, we generate 10,000 light curves using a simulated
$r$-band LSST cadence ($\sim$200 observations spread over 10 years;
Delgado et al.~2006). An example light 
curve with this cadence is shown in Figure~\ref{fig:LCegLSST}. 
Figure~\ref{fig:cadence} compares the resulting parameters for the
LSST cadence and a cadence of $\Delta t=10$~days over 10 years as in
the first test. The only significant change is an increased width for 
the best-fit $\hat{\sigma}$ distribution, where the accuracy of the
best-fit $\hat{\sigma}$ drops from 5\% to 13\%. 

\subsection{The Impact of Photometric Accuracy}
\begin{figure}[h!]
\epsscale{.61}
\plotone{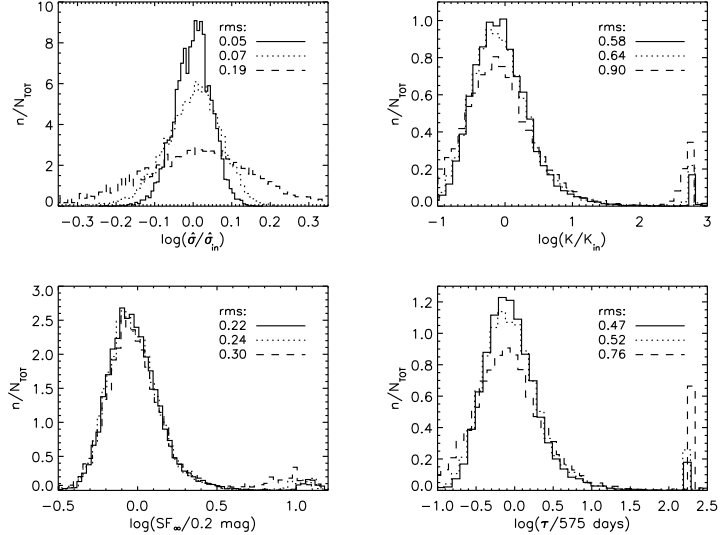}
\caption{\footnotesize Distributions of
  $\hat{\sigma}=\rm{SF}_{\infty}/\sqrt{\tau}$, $K=\tau\sqrt{\rm{SF}_{\infty}}$,
  $\tau$, and $\rm{SF}_{\infty}$ normalized by the input values for
  10,000 simulated light curves with a 10-year $r$-band LSST cadence
  and a photometric accuracy of 0.01 mag (solid), 0.03 mag (dotted), and
  0.1 mag (dashed). The rms of each distribution is listed in the legend.}
\label{fig:errors}
\end{figure}

As a third test, we evaluate the effect that the photometric accuracy
has on the fitted parameters by generating 10,000 light curves with
the $r$-band LSST cadence and photometric errors drawn from a Gaussian
of width $\sigma=0.03$ mag.  The same is done for $\sigma=0.1$ mag. The
output parameters for these two sets of light curves are compared to
those from the second test, where $\sigma=0.01$ mag, in
Figure~\ref{fig:errors}. As expected, the distributions become broader
as the photometric error bars increase and the fits become more
uncertain. While a photometric accuracy of  $\sigma=0.1$ mag leads to
a significant increase in the $\hat{\sigma}$ errors  
compared to the $\sigma=0.01$ mag case (a factor of 4 dex), an accuracy of
$\sigma=0.03$ mag leads to only a marginally wider distribution
(40\%). We conclude that a photometric accuracy of $\sigma=0.03$ mag
is sufficient to obtain well-constrained variability parameters for
typical quasars.  

\subsection{Selection Completeness for a Realistic LSST Simulation}
\begin{figure}[h!]
\epsscale{.5}
\plotone{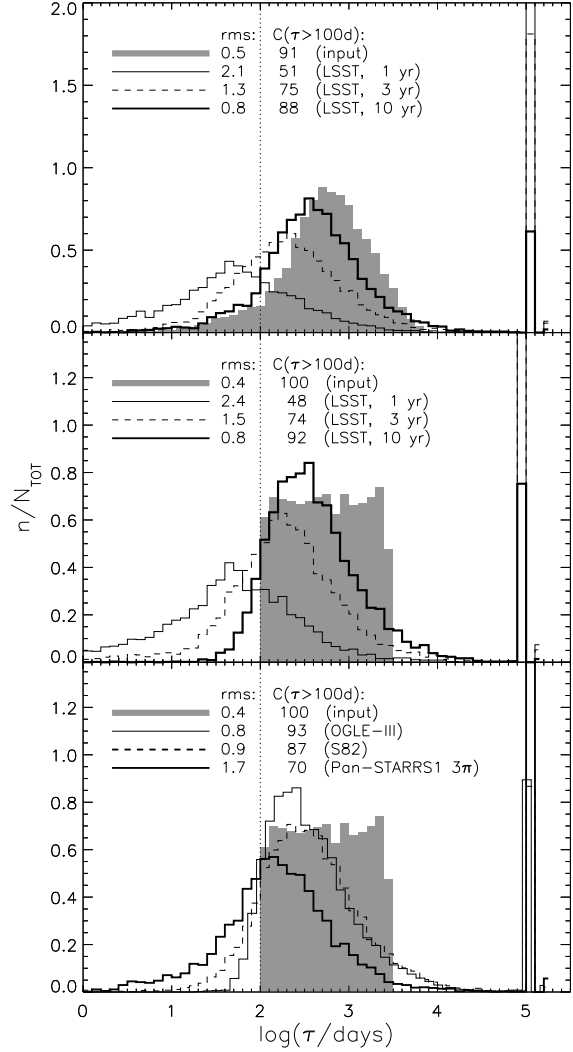}
\caption{\footnotesize  \emph{Top:} Distributions of  
  $\tau$ as compared to the input distribution (shaded histogram) for
  $\sim$7,000 simulated light curves with the $r$-band LSST cadence, a
  photometric accuracy of 0.03 mag, and lengths of 1 year (thin solid),
  3 years (dashed), and 10 years (thick solid).  The
  rms of each distribution is listed in the legend, as well as the
  completeness of the quasar selection criteria $\tau \ge 100$~days
  (dotted line). Note that light curves significantly longer
  than 1 year are required to achieve $>50$\% completeness.  
\emph{Middle:} As in top panel but for a uniform input distribution of
  $\tau$ within the range $100<\tau <10^{3.5}$~days.  
\emph{Bottom:} As in middle panel but for OGLE-III (thin solid), 
    S82 (dashed), and Pan-STARRS1~3$\pi$ (thick solid) cadences, which
    span 7, 10, and 3~years, respectively.  The denser time sampling
    of the OGLE-III survey leads to a higher completeness than S82. 
}
\label{fig:comp}
\end{figure}

As our final test, we estimate the completeness of quasar selection
based on the $\tau>100$~days criterion alone for three different light
curve lengths (1, 3, and 10 years), the $r$-band LSST cadence, and a
photometric accuracy of $\sigma=0.03$ mag (achieved at $i \approx 22$). 
Now, instead of using fixed input $\tau$ and $\rm{SF}_{\infty}$
values, we use the full observed $r$-band distributions for the S82
quasar sample as presented in MacLeod et al.~(2010). The distribution
of input $\tau$ is shown as the gray histogram in the top panel of
Figure~\ref{fig:comp}. The output distributions for each light curve
length are also shown.  We compute the completeness as the fraction of
the output distribution with a best-fit $\tau$ exceeding 100 days. The
completeness for the input distribution is 91\% (different than the
94\% reported in Table~\ref{tab:cuts} because all S82 quasars are
considered, without $\rm{rms}>0.05$ or $i<19$ mag limits). For light
curve lengths of 10, 3, and 1 year(s), $C$ is estimated to be 88\%,
75\%, and 51\%, respectively.  Notably, with 10 years of observations,
imperfectly measured $\tau$ causes only $\sim$3\% of the sample to
``leak'' below the $\tau=100$~day boundary. 
Because it is not known what the contamination will be
like for a typical LSST field, we cannot estimate the efficiency for
the $\tau>100$~days selection. With the aid of multi-band imaging,
colors can be used along with a relaxed $\tau$ cut to obtain highly
complete samples early in the survey ($\sim$1 year). 

While the light curve length clearly has an impact on the best-fit
time scale distribution, we also wish to estimate the effect of different
cadences in general. This is motivated by the fact that Koz{\l}10 found a
large fraction of quasars concentrated at smaller $\tau$ when applying
the DRW model to OGLE-II and OGLE-III light curves, which are more
densely-sampled than the SDSS S82 data.  Also, the
Pan-STARRS1~$3\pi$ survey will obtain 6 epochs in a given band over 3
years, as opposed to 60 epochs for LSST.  Therefore, we wish to 
determine whether a $\tau$ selection can still deliver decently 
complete samples of quasars in a typical 3~year-long Pan-STARRS survey.
We use Table~4 in Schm10 to simulate a typical Pan-STARRS cadence and
assume that the $griz$ bands can be combined to give 24 total 
epochs over 3 years (we omit the $Y$ band because of its lower
sensitivity). 

To make the comparison, we use a uniform input distribution of
$\log(\tau)$ and $\log(\rm{SF}_{\infty})$ within the region defined by  
$2<\log(\tau/\rm days)<3.5$ and $-1.4<\log(\rm{SF}_{\infty}/\rm mag)<-0.2$. 
We do not use the observed $\tau$ distribution for SDSS quasars for
the input, as in the previous test, in order to ignore any potential
effects of a biased $\tau$ distribution caused by the SDSS sampling.
We compare the fits at a fixed photometric uncertainty (0.03~mag),
therefore assuming that the absolute depth of each survey does not
play a role in these comparisons\footnote{\footnotesize Because the
  LSST survey has a fainter magnitude limit than OGLE, SDSS, or
  Pan-STARRS, it will have a higher completeness for quasar
  selection. Since we wish to compare differences due to cadencing
  only, we compare at a fixed photometric uncertainty. We note that a
  fixed uncertainty will induce some differences in the $\tau$
  distribution between surveys from what would realistically be
  observed.  However, that is a higher order question than what is
  examined here.}.  
In the middle and bottom panels of Figure~\ref{fig:comp}, we compare
the completeness results for the proposed LSST samplings to the 
typical 10~year-long S82, 7~year-long OGLE-III, and 3~year-long (24-epoch)
Pan-STARRS samplings. For each cadence, we compute the completeness as
the fraction of best-fit $\tau$ that does not ``leak'' below the
$\tau=100$~day boundary. The results show that the denser time
sampling of the OGLE-III survey leads to a higher completeness and a
more compact spread in output $\tau$ when compared to S82, and a
similar completeness when compared to the 10-year $r$-band LSST
cadence ($\sim$93\%).  We note that the time scales for the minority
of OGLE-III objects that also have OGLE-II light curves will be even
better constrained, since for these the total duration is
$\sim$10~years.  For Pan-STARRS1~$3\pi$, $C$ is estimated to be 70\%
for $\tau>100$~days after combining the four $griz$ bands.  This is
comparable to the result for a 3-year $r$-band LSST cadence. For a
single Pan-STARRS band (6~epochs over 3~years), $C$ will likely be significantly
worse, so in this case an alternate selection method (such as one
based on $\hat{\sigma}$) may be optimal in the absence of color selection. 

\subsubsection{Comparison to Butler \& Bloom (2010)}
\label{sec:compareBBMC}

We carried out similar simulations for the B\&B method, using the 
same uniform input distribution of $\log(\tau)$ and
$\log(\rm{SF}_{\infty})$.  Based on the mean trend of $\hat{\sigma}^2$ 
with g magnitude found for the S82 quasar sample in MacLeod et
al.~(2010), we assigned each light curve an apparent magnitude of 
\begin{equation}
  g = 0.334\log{\hat{\sigma}^2} + 21.38 + G(0.8),
\label{eq:mu}
\end{equation}
where $G(0.8)$ is a Gaussian random deviate of width 0.8 mag, 
matching the observed scatter about the mean trend. 
        
\begin{figure}[t!]
\epsscale{.8}
\plotone{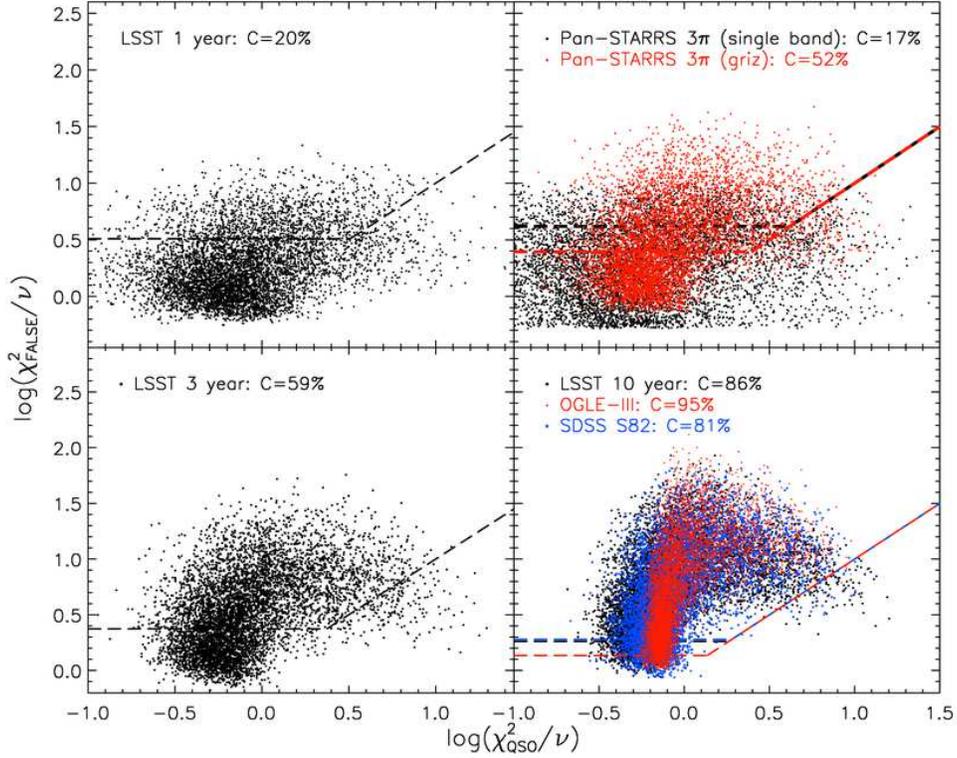}
\caption{\footnotesize  Distributions of $\chi^2_{\rm{FALSE}}/\nu$ and
  $\chi^2_{\rm{QSO}}/\nu$ resulting from fitting $\sim$7,000 simulated
  DRW light curves with various cadences (listed in the legends) using
  the B\&B technique.  The completeness of the quasar selection criteria 
  (the fraction above the dashed lines) for each cadence is listed.
}
\label{fig:compBB}
\end{figure}

Figure~\ref{fig:compBB} shows the output distributions of
$\chi^2_{\rm{FALSE}}/\nu$ and $\chi^2_{\rm{QSO}}/\nu$, as well as the
B\&B selection limits for a $5\sigma$ significance cut against false
positives. The figure legends list the cadences and the resulting
completeness $C$.  As survey length increases, $C$ increases because
of an overall increase in $\chi^2_{\rm{FALSE}}/\nu$. 
For the single-year LSST cadence, $C$ is only 20\%, and we find a
similar result for the 3-year single-band Pan-STARRS $3\pi$ cadence.
When combining the Pan-STARRS $griz$ bands, $C$ increases to 52\%, slightly lower
than the 3-year LSST cadence.  The LSST 10-year completeness is lower
than what was found when choosing $\tau \ge 100$~days in the previous
section (86\% vs.\ 92\%), but it could
be increased by weakening the cut on $\chi^2_{\rm{FALSE}}/\nu$. However, 
because the contamination rate is not known for LSST, it is unclear as
to how much this cut could be varied without a major decrease in
$E$.  Again, the OGLE cadence leads to the highest
$C$ due to the large number of epochs. We also find a
significant decrease in $C$ for the S82 cadence from what is observed
in the eleventh row of Table~\ref{tab:cuts} (81\% vs.\ 96\%). This
may be due to the prescription for generating mean
magnitudes. A larger $C$ is expected if we were to use the scaling in Table~1 in
B\&B, because the $\chi^2_{\rm{FALSE}}/\nu$ would likely be larger overall (i.e.,
white noise would be less likely). 

\clearpage

\section{Discussion and Summary}
\label{sec:summary}

We used the light curves for variable point sources
from SDSS Stripe 82 to determine how well quasars can be separated
from stars with the aid of photometric variability.  The Stripe 82
data set, designed to search for supernovae, is excellent for quasar
variability studies since it provides light curves for a large
statistical sample of objects to a faint limiting magnitude (see
Boutsia et al.~2010 for another example of using supernova surveys to
search for quasars). The data set used here contains precise
photometry (using recalibrated light curves by Ivezi\'{c} et al.~2007
and Sesar et al.~2007), as well as  
extensive spectroscopy, which has enabled us to develop an efficient
method for selecting quasars while maintaining high completeness.  
The variability model used here (a damped random walk, see KBS09; Koz{\l}10)
is robust enough to handle sparse or irregularly-sampled data.   
By modeling the variability as a stochastic process described by the
exponential covariance matrix $S_{ij} = \sigma^2\exp(-|t_i-t_j|/\tau)$, 
only O($N$) operations are required to determine the model parameters
for a light curve with $N$ data points.  This is an important algorithmic 
feature in the context of upcoming massive synoptic surveys, such as
LSST. 

The DRW model provides unique information on the
characteristic time scale, $\tau$, for which a range has been observed
for quasars (Sesar et al.~2006; KBS09; Koz{\l}10; MacLeod et al.~2010).   
Compared to selecting solely on the short-term behavior of
the $SF$ (the parameter $\hat{\sigma}$), the inclusion of $\tau$
information boosts $C$ from 90\% to 97\% while maintaining an
efficiency (purity, $E$) at 80\%, based on variability
alone. Conversely, for $C=98$\%, the $\tau$ information boosts $E$
from about 60\% to 75\%, while at $C=90$\%, $E$ rises from about 80\%
to 85\%. This method enables the selection of $z\sim 3$ quasars, where
color selection largely fails. With the aid of $u-g$ and $g-r$ color
selection, the efficiency can be further boosted to 96\% with a
reasonably high completeness of 93\%. It is worth reiterating that the
$E$ determined here is only a lower limit because the
spectroscopically confirmed quasar sample is not 100\% complete.  
This performance is accomplished simply by imposing a lower limit on
$\tau$ and an upper limit on $\hat{\sigma}$. 
The separation might improve when using more advanced
techniques such as kernel density estimation (e.g., Richards 2008)
rather than imposing simple limits.  

By using the parametrization $SF(\Delta t)=A(\Delta t/1 \hbox{yr})^{\gamma}$, 
Schm10 was able to select quasars from UVX objects with $E\simeq 95$\% at
90\% completeness when selecting on $A$ and $\gamma$. This selection
is similar to one using $\hat{\sigma}$ alone, and provides a
reasonable method for selecting quasars from short ($\lesssim 1$~year)
light curves. With longer light curves, the inclusion of $\tau$
information significantly improves the performance.  
The differences in $C$ and $E$ are summarized in the bottom panels of 
Figure~\ref{fig:optimaltausig} and illustrated in
Figure~\ref{fig:sftausig}.  If we restrict our initial sample to
UVX objects and simply use $\tau\geq 100$~days to select quasars, we
find that $C$ improves from the 90\% found by Schm10 
to 95\%, while $E$
improves from about 95\% to 97\%. When combining 
the selection criteria $\tau \geq 10^{1.6}$~days and $\hat{\sigma}
\leq 10^{-0.2}$ mag yr$^{-1/2}$, we find that $C$ improves further to 
98\% with $E=97\%$.  
We note that the DRW model has benefits beyond the improved selection
efficiency compared to Schm10. In addition to being a faster algorithm
[O($N$) rather than O($N^2$)],  
the DRW model also produces information on the physics of the
accretion disk since $\tau$ and 
$\rm{SF}_{\infty}$ are physical parameters that show correlations with
luminosity, black hole mass, and wavelength (see MacLeod et
al.~2010) and provides an approach for classifying other aperiodic sources 
as well (see Koz{\l}10). The latter two points are also advantages of our approach 
over that in B\&B.  For quasar selection, both methods yield similar results 
for completeness and efficiency. 

Our selection criteria identify 255 objects with $i<19$ and 
$-35\,^{\circ}<$~RA~$<50\,^{\circ}$ that vary on long time
scales ($\tau > 100$~days with $\Delta L_{\rm{noise}}>2$) and do not
presently have SDSS spectra; the majority of them are likely
quasars. Based on their light curves, several are expected to be
interesting long-time scale variable objects, such as AGB stars. We
plan to obtain follow-up spectroscopy for these objects
(Table~\ref{tab:targets}) with the goals of either confirming their
quasar nature or identifying the major contaminants in quasar selection based
on variability.    

The DRW provides a complete statistical model for quasar
variability. This is especially useful for evaluating the success
of quasar selection in upcoming surveys, and enables us to estimate
the completeness for different survey lengths, cadences, and
photometric accuracy.  We cannot estimate $E$, however, since it is
dependent on an unknown makeup of the faint variable optical sky. For
a typical (simulated) LSST cadence over 10 years and a photometric
accuracy of 0.03 mag (achieved at $i \approx 22$), $C$ is expected to
be 88\% for a sample selection criterion of $\tau>100$~days.   For
typical variable quasars, the best-fit distributions of $\tau$ and
$\rm{SF}_{\infty}$ are biased for surveys shorter than 10 years (by about
70\% and 40\%, respectively, for a length of 3 years). However,
$\hat{\sigma}$ is well-constrained, and with the aid of multi-band
imaging, colors can be used along with a relaxed $\tau$ cut to obtain
highly complete samples early in the LSST survey ($\sim$1 year).  
For the Pan-STARRS1~$3\pi$ survey, $C$ is estimated to be 70\%
for a $\tau>100$~day selection after combining the $griz$ bands, which is
comparable to the result for a 3-year $r$-band LSST cadence. 
The dense cadence of the LSST will provide tighter constraints on
$\tau$ for a survey length of 10 years.  This improvement is also important for
relating observations to the physics of accretion disks (see MacLeod et al.~2010,
and references therein). The most important conclusion of this work is that,
given an adequate survey cadence, photometric variability provides an even better 
method than color selection for separating quasars from stars.

\acknowledgments

We acknowledge support by NSF grant AST-0807500 to the University of
Washington, and NSF grant AST-0551161 to LSST for design and development activity.
CSK and SK acknowledge support by NSF grants AST-0708082 and
AST-1009756. RRG gratefully acknowledges support from NASA {\it  
Chandra} grants AR9-0015X and AR0-11014X.

    Funding for the SDSS and SDSS-II has been provided by the Alfred P.\ Sloan Foundation, the Participating Institutions, the National Science Foundation, the U.S.\ Department of Energy, the National Aeronautics and Space Administration, the Japanese Monbukagakusho, the Max Planck Society, and the Higher Education Funding Council for England. The SDSS Web Site is http://www.sdss.org/.

    The SDSS is managed by the Astrophysical Research Consortium for the Participating Institutions. The Participating Institutions are the American Museum of Natural History, Astrophysical Institute Potsdam, University of Basel, University of Cambridge, Case Western Reserve University, University of Chicago, Drexel University, Fermilab, the Institute for Advanced Study, the Japan Participation Group, Johns Hopkins University, the Joint Institute for Nuclear Astrophysics, the Kavli Institute for Particle Astrophysics and Cosmology, the Korean Scientist Group, the Chinese Academy of Sciences (LAMOST), Los Alamos National Laboratory, the Max-Planck-Institute for Astronomy (MPIA), the Max-Planck-Institute for Astrophysics (MPA), New Mexico State University, Ohio State University, University of Pittsburgh, University of Portsmouth, Princeton University, the United States Naval Observatory, and the University of Washington.

\end{document}